\documentclass[aps,prl,twocolumn,superscriptaddress,longbibliography,nofootinbib]{revtex4-2}

\usepackage{lmodern} 
\usepackage[utf8]{inputenc} 
\usepackage[T1]{fontenc} 
\usepackage{microtype} 
\usepackage[english]{babel}
\usepackage{color}
\usepackage{setspace}
\usepackage{graphicx}
\usepackage{amsmath}
\usepackage{amssymb}
\usepackage{slashed}
\usepackage{cancel}
\usepackage[bbgreekl]{mathbbol}
\usepackage[usenames,dvipsnames]{xcolor}
\usepackage[colorlinks=true, linkcolor=black, citecolor=ForestGreen]{hyperref}
\usepackage{changes}

\usepackage[mathscr]{euscript}

\begin{document}

\begin{flushright}
CPHT-RR052.072024
\end{flushright}

\title{Linear dynamical stability and the laws of thermodynamics}
\author{Blaise Gout\'eraux}
\email{blaise.gouteraux@polytechnique.edu}
\affiliation{CPHT, CNRS, \'Ecole polytechnique, Institut Polytechnique de Paris, 91120 Palaiseau, France}
\author{Eric Mefford}
\email{emeffordphysics@gmail.com}
\affiliation{Department of Physics and Astronomy, University of Victoria, Victoria, BC V8W 3P6, Canada}

\date{\today}

\begin{abstract}
We show that the dynamical stability under linear perturbations of interacting systems in the hydrodynamic regime follows from the first and the second laws of thermodynamics. Our argument extends to systems with spontaneously or softly broken symmetries and in the presence of magnetic fields.
\end{abstract}
\maketitle

\section{Introduction}

Thermodynamics is the universal framework to describe equilibria of physical systems at finite temperature. Deviations away from thermal equilibrium present a much greater challenge. In the regime where the scales of variation in time and space are large compared to those where local thermal equilibrium is established, the effective theory governing these deviations is hydrodynamics \cite{landaubook,chaikinlubensky1995}, and its extensions when symmetries are spontaneously or softly broken. More precisely, hydrodynamics determines the dynamics of conserved charges after integrating out microscopic, quickly-relaxing degrees of freedom. 

Spatial fluxes are associated to the densities of thermal expectation values of conserved operators, and characterize the flow of the densities outside of each infinitesimal volume of the system. These are generally not conserved operators themselves, and obtain so-called constitutive relations: their expectation values in the local equilibrium state can be written as local expansions in gradients of the densities, parameterized through a number of transport coefficients. 
The form that these gradient corrections can take is constrained by the physical requirement that the divergence of the entropy current is positive-definite, i.e. that entropy can only be created and not destroyed. This imposes equality-type constraints on `non-hydrostatic, non-dissipative' transport coefficients, that are out-of-equilibrium corrections which do not contribute to entropy production, or inequality-type constraints on `non-hydrostatic dissipative' transport coefficients, which do, \cite{Bhattacharyya:2013lha,Bhattacharyya:2014bha,Haehl:2014zda,Haehl:2015pja,Jain:2018jxj}. Besides these, `hydrostatic' gradient corrections capture small deviations from spatial homogeneity in the static partition function and by definition do not produce entropy.

In the thermal equilibrium state, the local stability of the system is determined by the positive-definiteness of the static susceptibility matrix, i.e. of second derivatives with respect to thermodynamic sources of the static partition function. Away from equilibrium, the system is stable under linear perturbations if the retarded Green's functions do not display physical poles in the upper half complex frequency plane: poles with a positive imaginary part $\Im \omega > 0$ lead to exponential growth of linear perturbations, and so trigger instabilities.\footnote{Depending on the choice of hydrodynamic frame, that is, on the definition of thermodynamic sources away from equilibrium, there can be spurious, unphysical poles in the upper half plane. These poles lie outside the hydrodynamic regime and do not signal instabilities of physical theories with a hydrodynamic limit. They are important only when hydrodynamics is promoted to a UV-complete theory. In this case, special frames can be chosen so that such poles are absent, \cite{Kovtun:2019hdm,Bemfica:2020zjp}.} The dispersion relations of such poles, i.e. their complex frequency expressed as an expansion in powers of the (real-valued) wavevector in Fourier space, depends in a complicated way on the thermodynamics of the state and on the coefficients of derivative corrections. 

In this work, we show that local thermodynamic stability and positivity of entropy production are sufficient to ensure linear dynamical stability of fluids. While this statement can be made on a case-by-case basis in sufficiently simple situations with a small number of transport coefficients, we are not aware of a general proof of it.\footnote{There exists a large body of work explicating the relation between stability of thermodynamic equilibrium, positivity of entropy production, dynamical stability and causality in the context of relativistic hydrodynamics, \cite{Hiscock:1983zz,Hiscock:1985zz,Geroch:1990bw,Geroch:1991}, and more recently \cite{Gavassino:2021cli,Gavassino:2021kjm,Gavassino:2021owo}. Here we do not assume relativistic boost invariance.} Showing it by direct comparison between the inequality-type constraints on transport coefficients stemming from positivity of entropy production and the constraint that hydrodynamic poles lie in the lower-half plane becomes a significant challenge when the number of transport coefficients proliferates, away from highly-symmetric systems (see e.g. analyses for boost-agnostic fluids, \cite{Novak:2019wqg,Armas:2020mpr}). Instead, our proof applies irrespective of invariance under any kind of boosts, and extends to cases where some of the global symmetries are spontaneously or softly broken, or in the presence of weak external gauge fields.\footnote{The global symmetries can be unconventional as in the kinematically constrained systems in \cite{Jain:2024kri}, where the observed instability at finite velocity can be traced back to a non-positive-definite susceptibility matrix.} 

We expect that our arguments can be extended to many other systems of interest with a hydrodynamic regime, such as hydrodynamics with chiral anomalies \cite{Son:2009tf}, QCD in the chiral limit \cite{Son:1999pa,Grossi:2020ezz,Grossi:2021gqi}, topological defects \cite{PhysRevB.42.9938,Davison:2016hno,Delacretaz:2017zxd,Armas:2022vpf,Armas:2023tyx}, spinful systems \cite{Gallegos:2021bzp,Gallegos:2022jow}, dipole symmetries \cite{Stahl:2023prt,Jain:2023nbf,Armas:2023ouk,Jain:2024kri,Glodkowski:2024ova}, systems with Carrollian symmetries \cite{Armas:2023dcz}, and magnetohydrodynamics in its higher-form symmetry formulation \cite{Grozdanov:2016tdf,Armas:2018zbe,Armas:2018atq}. 

On the other hand, phases in the vicinity of a continuous phase transition fall outside our arguments. In theories with a gravitational dual, the correlated stability conjecture of black branes \cite{Gubser:2000ec,Gubser:2000mm,Buchel:2005nt} was extensively investigated, and was reported not to apply in such cases as originally formulated \cite{Friess:2005zp,Buchel:2010wk}. New light degrees of freedom need to be included and treated in the theory of dynamical phase transitions \cite{Hohenberg:1977ym}. However, in theories with a large number of degrees of freedom, such as those with a gravitational dual, stochastic and quantum fluctuations are suppressed \cite{Kovtun:2003vj}. As we show in the Appendix for the case of a holographic superfluid, the susceptibility for the order parameter to acquire a non-zero expectation value indeed becomes negative for temperatures below the critical temperature, indicating that the symmetry broken phase is thermodynamically preferred---in sync with the results of the main text, this is mirrored by a low-energy mode with $\Im \omega > 0$. This suggests that our arguments can be extended to holographic continuous phase transitions, accounting for the dynamics of the amplitude mode of the order parameter, \cite{Herzog:2010vz,Donos:2022xfd,Donos:2022qao,Donos:2023ibv}

Throughout, we use the Einstein summation convention both on capital and smaller case Latin indices and raise or lower both with the Kronecker symbol.

\section{Linear stability of hydrodynamics}

The conservation equations take the general form:
\begin{equation}
\label{eoms}
\partial_t \langle O^A\rangle+\partial_i \langle j^{Ai}\rangle=f^A
\end{equation}
Angular brackets denote expectation values taken in a thermal ensemble at temperature $T$, $\langle \cdot\rangle=Tr\left(\cdot e^{-H/T}\right)$ where $H$ is the Hamiltonian of the system. The $O^A$ are the set of $N$ densities of conserved (slowly-relaxing) charges of the system, following from the global (approximate) symmetries, and $A$ is an index running over the entire set. For example, in the case of a fluid charged under a global $U(1)$ symmetry, $O^A={\epsilon,n,g^i}$ are the energy, charge and momentum densities and $N=d+2$, where $d$ is the number of spatial dimensions. The $j^{Ai}={j_\epsilon^i,j_n^i,\tau^{ij}}$ are the associated fluxes, with the $i$ index running over all spatial dimensions. To these densities are associated a set of sources $s^A$, e.g. in the case of a charged fluid, $s^A=\left(-1/T,\mu/T,v^i/T\right)$ with $\mu$ the chemical potential and $v^i$ the fluid velocity. 

Finally, the $f^A$ are possible inhomogeneous terms which account for the weak breaking of the conservation laws due to external fields.\footnote{The external fields can be background gauge fields associated with local symmetry transformations or external fields which explicitly break the symmetry, and are appropriately incorporated into definitions of the hydrodynamic sources $s_A$.} The existence of a symmetry-breaking scale means that we need to employ a double-expansion scheme: the standard expansion in gradients, controlled by the hierarchy of scales between the spatial and temporal variations of the fluxes and densities compared to the local equilibration scales, $\tau_{eq}\partial_t\ll1$, $\ell_{eq}\partial_i\ll1$; and an expansion in the symmetry-breaking scales $\tau_{eb}$, $\ell_{eb}$, which we also take to be large compared to the scales at which local equilibrium is established,  $(\tau_{eb},\ell_{eb})\gg(\tau_{eq},\ell_{eq})$ but with no special hierarchy compared to gradients.

Our work is concerned with the spectrum of the linearized equations of motion. To arrive at these, we expand the sources and densities about their equilibrium values and treating individual fluctuations as plane waves,\footnote{In order to expand in plane waves, the linearized equations of motion cannot have any dependence on the spatiotemporal coordinates beyond derivatives with respect to these coordinates, so any spatial dependence must drop out the linearized constitutive relations or cancel against similar terms from other parts of the equations.} $s^A = \bar{s}^A + \delta s^{A}(q,\ell_{eb}) e^{-i\omega t + iq_j x^j}$ and $O^A = \bar{O}^A  +\delta O^A(q,\ell_{eb}) e^{-i\omega t+iq_j x^j}$. In the main text, we will assume $\partial_i \bar{s}_A  = 0$  for ease of reading, though in the Appendix we show that our results continue to hold when this is not the case. 

The main assumption of hydrodynamics is that we can express the $\delta O^A$, $\delta j^{Ai}$ and $\delta f^{A}$ in a small $|q|$ expansion in terms of the fluctuating sources $\delta s^A$, with $q = \sqrt{q_i q^i}$. A particularly important instance of these `constitutive relations' is 
\begin{align}
\delta O^A = \chi^{AB}(q,\ell_{eb})\delta s_B = (\chi_0^{AB}+i\chi_1^{ABi}q_i+...)\delta s_B
\end{align}
where $\chi^{AB}(q, \ell_{eb})$ is the static susceptibility matrix. In the Appendix, we will discuss this expansion in terms of more specific constraints on the hydrodynamic dispersion relations. The matrices $\chi_{0,1}$ can depend on $\ell_{eb}$ in two ways: either through hydrostatic terms allowed in the presence of symmetry-breaking sources, which after integrating out the sources simply redefine thermodynamic quantities; or through out-of-equilibrium terms `masquerading' as hydrostatic terms after integrating out the sources -- see the discussion of approximate symmetries below for concrete examples. Here, we emphasize that the hydrodynamic expansion is only expected to be justified up to some radius of convergence $|q_c|\geq 0$, \cite{Withers:2018srf,Grozdanov:2019kge,Grozdanov:2019uhi,Heller:2020uuy,Heller:2020hnq,Heller:2021yjh}. Instabilities may arise from conservation equations for large $|q|$ \cite{Kovtun:2019hdm}, but we do not expect that these are physical, and we will not try to apply our argument to such cases. 

After expanding to linear order, the equations of motion read,
\begin{align}
\label{linearized_eom}
\left(i[q\bar{v}-\omega(q,\ell_{eb})] \delta^{AB}+\tilde{M}^{AB}(q,\ell_{eb})\right)\delta s_B(q, \ell_{eb}) = 0
\end{align}
where we have explicitly written a frame dragging term originating from a possible nonzero fluid velocity in equilibrium. The $\tilde{M}^{AB}(q,\ell_{eb})$ follow from consistency conditions of the constitutive relations with a local version of the first law of thermodynamics, leading to hydrostatic constitutive relations, and a local version of the second law of thermodynamics, leading to non-hydrostatic constitutive relations. The hydrostatic constitutive relations are also consistent with the second law in that they do not lead to entropy production. For ease of reading, we have relegated a general derivation of this matrix to the Appendix. 

The result of these consistency conditions is that the matrix appearing in the linearized equations of motion takes a generic form
\begin{align}
\label{Mform}
\begin{split}
M^{AB} &= -\ell_{eb}Q^{[AB]} + iq N^{(AB)} +\ell_{eb}\tilde{Q}^{AB}+q^2\tilde{N}^{AB},\\
\tilde{M}^{AB} &= (\chi^{-1}(q,\ell_{eb}))^{AC}M^{CB}(q,\ell_{eb})
\end{split}
\end{align}
The linearized equations of motion are seen to project onto components parallel to the wavevector, i.e. $\bar{v} \equiv \hat{q}\cdot \langle v^i \rangle$ with $\hat{q}^i = q^i/q$. Furthermore, $Q, \tilde{Q}, N,$ and $\tilde{N}$ are all real matrices which are functions of $q$ and $\ell_{eb}$ and to leading order in the hydrodynamic expansion are constant matrices. In other words, the non-tilded parts of $M^{AB}$, which come from the hydrostatic constitutive relations, are given by $i$ times a Hermitian matrix. The tilded terms of $M^{AB}$, which come from the non-hydrostatic constitutive relations, can have any overall symmetry under $A\leftrightarrow B$, but the second law requires that symmetric pieces must be positive definite,
\begin{align}
\partial_ts + \partial_ij_s^i = \frac{\Delta}{T} \geq 0 
\end{align}
with
\begin{equation}
 \frac{\Delta}{T}= \ell_{eb}\delta s_A\tilde{Q}^{(AB)}\ell_{eb}\delta s_B+q \delta s_A\tilde{N}^{(AB)}q\delta s_B \,.
\end{equation}
The form of $\tilde{M}^{AB}$ is sufficient to imply stability of the hydrodynamic modes, meaning $\Im \omega \leq 0$, whenever $\chi^{AB}(q, \ell_{eb})$ is a positive definite matrix, as we now show.

The operator $M^{AB}$ can be decomposed into real and imaginary components $M^{AB} = M^{AB}_{\text{re}} + i M^{AB}_{\text{im}}$ where $M^{AB}_{\text{re}}$ and $M^{AB}_{\text{im}}$ are matrices with purely real entries. Furthermore, from \eqref{Mform}, $M^{(AB)}_{\text{re}}$ is constrained to be positive definite and $M^{[AB]}_{\text{im}} = 0$. These facts tell us that the $N$ eigenvalues of $M^{AB}$, written as $m^{(A)}$ for $A=1,...,N$ have a general form $m^{(A)} = \alpha^{(A)} + i\beta^{(A)}$ with $\alpha^{(A)} \geq 0$. To see this, write the corresponding eigenvector as $\delta s^{(A)}_{B} = u^{(A)}_B+iv^{(A)}_B$. Then matching real and imaginary parts of the eigenvalue equation $M^{CB}\delta s^{(A)}_B = m^{(A)}\delta s^{(A)}_C$ implies
\begin{align}
\begin{split}
M^{CB}_{\text{re}}u^{(A)}_B - M^{CB}_{\text{im}}v^{(A)}_B &= \alpha^{(A)}u^{(A)}_C - \beta^{(A)} v^{(A)}_C \,,\\
M^{CB}_{\text{re}}v^{(A)}_{B} + M^{CB}_{\text{im}}u^{(A)}_{B} &= \alpha^{(A)}v^{(A)}_C + \beta^{(A)} u^{(A)}_C
\end{split}
\end{align}
and positivity requires that $x_CM^{(CB)}_{\text{re}}x_B \geq 0$ for all vectors $x_C$, including $v^{(A)}$ and $u^{(A)}$. So,
\begin{align}
\begin{split}
0&\leq u^{(A)}_CM^{(CB)}_{\text{re}}u^{(A)}_{B} + v^{(A)}_CM^{(CB)}_{\text{re}}v^{(A)}_{B} \\
&= 2\cancel{u^{(A)}_{C}M^{[CB]}_{\text{im}}v^{(A)}_{B}} + \alpha^{(A)} (u^{(A)}_C u^{(A)}_{C}+  v^{(A)}_C v^{(A)}_{C})
\end{split}
\end{align}
implying that $\alpha^{(A)}\geq 0$. 

When $\chi^{AB}(q, \ell_{eb})$ is a Hermitian positive-definite matrix, we can define a Hermitian positive-definite matrix $\tilde{\chi}^{AB} \equiv (\sqrt{\chi})^{AB}$, a vector $\delta \tilde{s}_A = \tilde{\chi}^{AB}\delta s_B$, and a matrix $\dot{M}^{AB} \equiv (\tilde{\chi}^{-1}\cdot M \cdot \tilde{\chi}^{-1})^{AB}$. In terms of this vector, the linearized equations of motion read
\begin{align}
\left(-i[\omega(q)-q\bar{v}]\delta^{AB} + \dot{M}^{AB}\right)\delta \tilde{s}_B = 0\,.
\end{align}
This is a similarity transform, so the eigenvalues of $\dot{M}^{AB}$ are the same as those of $\tilde{M}^{AB}$. Since $\tilde{\chi}$ generically has complex entries, we establish positive-definiteness of the Hermitian part of $\dot{M}^{AB}$, given by
\begin{align}
\begin{split}
\frac{1}{2}(\dot{M}+\dot{M}^\dagger)^{AB} &= \frac{1}{2}(\tilde{\chi}^{-1}\cdot (M+ M^\dagger) \tilde{\chi}^{-1})^{AB} \\
&= (\tilde{\chi}^{-1}\cdot M_{\text{re, sym}}\cdot \tilde{\chi}^{-1})^{AB}
\end{split}
\end{align}
where $M_{\text{re, sym}} = M^{(AB)}_{\text{re}}$. Finally, for any vector $x^A \in \mathbb{C}^N$, we may define $(\tilde{\chi}^{-1})^{AB}x^B = y^A + i z^A$ and $(x^A[\tilde{\chi}^{-1}]^{AB})^* = y^B - i z^B$ where $y^A \in \mathbb{R}^N$ and $z^A\in \mathbb{R}^N$. Then,
\begin{align}
 \frac{1}{2}(x^*)^A(\dot{M}+\dot{M}^\dagger)^{AB}x^B = y^A M_{\text{re}}^{(AB)} y^B \geq 0
 \end{align}
 where the last inequality follows from positivity of the symmetric part of $M_{\text{re}}$. Hence, the Hermitian part of $\dot{M}^{AB}$ is positive definite. Following similar arguments as earlier for the matrix $M$, then, writing the eigenvalues of $\tilde{M}^{AB}$ as $\tilde{m}^{(A)} = \tilde{\alpha}^{(A)} + i \tilde{\beta}^{(A)}$, where $\tilde{\alpha}^{(A)}$ and $\tilde{\beta}^{(A)}$ are real, positivity of the Hermitian part of $\dot{M}^{(AB)}$ implies positivity of $\tilde{\alpha}^{(A)}$.
 
We have therefore shown that, when the system is thermodynamically stable, the dispersion relations of the hydrodynamic modes, 
\begin{align}
\omega^{(A)} = q\bar{v} - i\tilde{\alpha}^{(A)}+\tilde{\beta}^{(A)} \,,
\end{align}
have $\Im \omega^{(A)} = -\tilde{\alpha}^{(A)} \leq 0$. On the other hand, when the system is not thermodynamically stable, $\chi^{AB}$ is not positive definite and $(\sqrt{\chi})^{AB}$ is therefore not Hermitian, so the argument fails. 

In other words, sufficient conditions to guarantee that hydrodynamic theories are linearly stable are
\begin{enumerate}
\item The matrix $M^{AB}$ takes the form in \eqref{Mform};
\item The matrix $\chi^{AB}$ is positive definite.
\end{enumerate}
 
We now proceed to give a few examples to illustrate our results.

\subsection{Fluids charged under a global U(1)}

The simplest example is a fluid charged under a global U(1), with $s^A = \{-1/T,\mu/T, v^i/T\}$ and $O^A = \{\epsilon, n, g^i\}$, respectively. We set $\ell_{eb}=0$. The pressure is defined as $p\equiv T s+\mu n+v_i g^i-\epsilon$ and the first law reads $ds= s^A dO_A = (-1/T)d\epsilon+(\mu/T) dn+(v_{i}/T)dg^i$. The constitutive relations for the fluxes are
\begin{equation}
\begin{split}
&j_n^{i}=nv^i+\tilde j_n^i\,,\quad j_\epsilon^{i}=(\epsilon+p) v^i+\tilde j_\epsilon^{i}\,,\\
&g^i=\rho v^i\,,\quad  \tau^{ij}=p\delta^{ij}+v^i g^j+\tilde \tau^{ij}
\end{split}
\label{constitutivechargedfluid}
\end{equation}
where $\rho$ is the momentum static susceptibility and tilded fluxes contain gradient corrections. The first law can be combined with the equations of motion, $\Delta/T-\partial_ij_s^i = \partial_t s = s^A\partial_t O^A= -s^a\partial_i j^{Ai}$, leading to the `adiabaticity equation,'
\begin{align}
\label{Deltaeq}
\partial_i(j_s^i + s_Aj^{Ai}) - j^{Ai}\partial_is_A = \Delta/T \geq 0.
\end{align}
It is easily checked that for $j_s^i = -s_Aj^{Ai} + v^i p/T$, the second law reads
\begin{equation}
\label{DeltaChargedFluids}
\frac{\Delta_0}{T}=-\tilde j_\epsilon^i\partial_i \left(-\frac{1}{T}\right)-\tilde j_n^i\partial_i\left(\frac{\mu}{T}\right)-\tilde\tau^{ij}\partial_i\left(\frac{v_j}{T}\right)\,.
\end{equation}
In other words, the hydrostatic (non-tilded) constitutive relations do not produce entropy (they lead to $\Delta = 0$), while the non-hydrostatic pieces may. Expanding around a state with $\partial_i\langle s_A\rangle = 0$, we may write $\delta \tilde{j}^{Ai} =\tilde{N}^{AB} iq\delta s_B$ in which case it is clear that the second law requires $\tilde{N}^{(AB)}$ to be a positive definite matrix, consistent with \eqref{Mform}. The antisymmetric part of $\tilde{N}^{AB}$ leads to $\Delta = 0$. The hydrostatic constitutive relations appear in \eqref{Mform} through
\begin{align}
N^{AB} &=2\bar{T}\left(p\delta^{\epsilon(\alpha}+\bar{O}^{(\alpha}\right)\left(\delta^{\beta)g}+\bar{v}\delta^{\beta)\epsilon}\right) 
\end{align}
which is symmetric, as advertised.

For a boost-invariant charged fluid, \cite{chaikinlubensky1995,Kovtun:2012rj}, there are three independent dissipative coefficients entering in $\tilde{N}$, and so stability of the hydrodynamic modes given positivity of entropy production and thermodynamic stability is easily verified.
However, in a charged fluid without boost invariance, due to the need to account for explicit dependence on the fluid velocity in the tensor structures allowed at first order in gradients, the number of non-hydrostatic coefficients increases to $26$, \cite{Novak:2019wqg,Armas:2020mpr}. Matching the constraints from linear stability and positivity of entropy production is significantly more involved. The spectrum consists of 2 sound modes, a mixed charge-heat diffusion mode, and $d-1$ shear diffusion modes,
\begin{align}
\begin{split}
&\omega_{s,\pm} = v_{s,\pm} q - i\Gamma_{s,\pm}q^2+\mathcal{O}(q^3), \\
&\omega_{n} = -iD_{n}q^2+\mathcal{O}(q^3), 
\;\; \omega_{sh} = -iD_{sh}q^2+\mathcal{O}(q^3)
\end{split}
\end{align}
In the absence of a thermodynamic instability, $\Gamma_{s,\pm} \geq 0, D_n \geq 0,$ and $D_{sh} \geq 0$ and the fluid is linearly dynamically stable.

\section{Spontaneous symmetry breaking}
If a global symmetry is spontaneously broken, then we need to include the gradients of the $M$ Goldstone fields in the $O^A$, $A=1\dots N+M$, \cite{chaikinlubensky1995}.  We use lower Greek indices $\beta=1\dots N$ to run over the densities of conserved charges and $a=1\dots M$, $M\leq N$ to enumerate the Goldstones, which we denote collectively $\partial_i\varphi^a$.\footnote{Here we set aside subtleties related to the counting of the number of Goldstone modes when spacetime symmetries are spontaneously broken by only considering the broken symmetry generators giving rise to independent Goldstone fields: e.g. we do not include broken rotation generators in the case of broken spatial translations. Then the Goldstone fields can indeed be labeled by indices $a$, which are running only over the set of independent broken generators.} Examples are the superfluid velocity $v_s^i=\partial_i\varphi$ for a superfluid, the strains $u_{ij}=\partial_{(i}u_{j)}+(1/2)\partial_i u^k\partial_j u^k$ for a crystalline solid, etc. Goldstones only appear through gradients due to a residual shift symmetry, which is the nonlinear realization of the spontaneously broken symmetry. This implies that we formally treat $\partial\varphi^a\sim\mathcal O(\partial^{0})$. The Goldstone fields obey `Josephson' evolution equations:
\begin{equation}
\frac{\partial\varphi^a}{\partial t}+X^a =0\,,\quad X^a=v^i\partial_i\varphi^a+Tg^a_\beta s^\beta+\tilde X^a\,,
\end{equation}
where $g^a_\beta s^\beta$ returns the source conjugate to the density of the broken symmetry, i.e. the matrix $g^a_{\beta}$ is a constant with no dependence on the sources.\footnote{For instance, for superfluids $s^\varphi=\mu$ while for crystals $s^{u^i}=v^i$.} $\tilde X^a$ stands for the dissipative corrections, which have been worked out in a number of different cases with or without boost invariance \cite{chaikinlubensky1995,Bhattacharya:2011eea,Bhattacharya:2011tra,Delacretaz:2017zxd,Armas:2019sbe,Armas:2020bmo,Armas:2022vpf}. It is useful to take a derivative of the Josephson equation
\begin{align}
\partial_t (\partial_j\varphi^a) + \partial_i (X^{a}\delta^{ji}) = 0
\end{align}
in which case this has the same `divergence-type' form as the equations for the other conserved densities.

The first law is modified as $d(p/T)\mapsto d(p/T)-(h^{ai}/T)d(\partial_i\varphi_a)$, defining the source $h^{ai}/T$ conjugate to $\partial_i\varphi^a$. The previous arguments can then be repeated, with suitable modifications to the constitutive relations. We now get 
$s_A j^{Ai}=-j_s^{i}+(p/T) v^i+(v^j\partial_j\varphi^a+T g^a_\beta s^\beta) (h^{ai}/T)$ and 
the constitutive relations are modified to
\begin{align}
\begin{split}
j^{\alpha i} &= (p\delta^{ij}+h^{ai}\partial_j\varphi^a)\,\delta^{\alpha g_j}\\
&\quad+\left(pv^i+h^{ai}\left[v^j\partial_j\varphi^a+Tg_\beta^{a}s^{\beta}\right]\right)\delta^{\alpha \epsilon}\\
&\quad+ O^\alpha v^i+h^{ai}g^a_{\alpha} + \tilde{j}^{\alpha i}
\end{split}
\end{align}
which in turn updates the $N$ matrix to
\begin{align}
\begin{split}
N^{\alpha\beta} &=\bar{T}\left(2h^ag^{a}_{(\alpha} + h^a[\bar{v}^j\partial_j\bar{\phi}^a+\bar{T}g^{a}_{\gamma}\bar{s}^\gamma]\delta^{\epsilon(\alpha}\right)\delta^{\beta)\epsilon} \\
&\quad+ 2\bar{T}\left(p\delta^{\epsilon(\alpha}+\bar{O}^{(\alpha}\right)\left(\delta^{\beta)g}+\bar{v}\delta^{\beta)\epsilon}\right) \\
&\quad+ \bar{T} h^a (\hat{q}^j\partial_j\bar{\phi}^a)\delta^{\epsilon(\alpha}\delta^{\beta)g}\\ 
N^{\beta a} &=N^{a\beta}= \bar{T}g^{a}_{\alpha}\left(\delta^{\alpha \beta} + \bar{T}\bar{s}^\alpha \delta^{\beta\epsilon}\right)\\
&\quad\quad\quad\quad +\bar{T}\partial_j\bar{\phi}^a \left(\delta^{\beta g_j}+\bar{v}^j\delta^{\beta\epsilon}\right),
\end{split}
\end{align}
where $h^a = \hat{q}_ih^{ai}$ and $ N^{ab}=0$. Importantly, the matrix $N^{AB}$ remains symmetric with only longitudinal components contributing to the dispersion relations. For instance, in the absence of explicit symmetry breaking, $\tilde{f}^A = 0$, and defining $\delta s^A = (\delta s_{\alpha}, \delta(h_a/T))^A$,
\begin{align}
\frac{\Delta}T = -\tilde{j}^{\alpha i}\partial_i s_\alpha - \tilde{X}^a\partial_i \left(\frac{h_{ai}}{T}\right)= q^2 \delta s^A\tilde{N}^{AB}\delta s^B,
\end{align}
with $\tilde{N}^{(AB)}$ a positive semi-definite matrix. It is useful to consider as explicit examples the cases of superfluids (spontaneous breaking of a global U(1) symmetry) and crystalline solids (spontaneously broken translations).

\subsection{\label{sec:superfluids} Superfluids}

In the case of superfluids, the first law becomes \cite{chaikinlubensky1995}
\begin{equation}
\label{1stlawU1}
d\epsilon=Tds+\mu dn+v_{ni}dg^i+h_i dv_{s}^i\,.
\end{equation}
where the normal velocity $v_n^i$ appears together with the superfluid source $h^i\equiv\delta f/\delta v_s^i= n_s v_s^i-\nu v_n^i$ defined from the free energy density $f$. $n_s$ is the superfluid density which quantifies the cost of superfluid velocity fluctuations in the free energy, while $\nu$ is a thermodynamic parameter reducing to $\nu=n_s$ in the presence of Galilean or Lorentz boost symmetry, \cite{Armas:2023ouk}. The equations of motion are supplemented by the Josephson relation
\begin{equation}
\partial_t v_s^i+\partial_i X=0\,,\quad X=v_n\cdot v_s+\mu+\tilde X
\end{equation}
together with the constitutive relations
\begin{equation}
\begin{split}
& j_n^{i}=nv_n^i+h^i+\tilde j_n^i\,,\quad g^i=\rho v_n^i+\frac{\mathcal{N}}{n_s}h^i\,,\\
& \tau^{ij}=p\delta^{ij}+v_n^i g^j+h^i v_s^j+\tilde \tau^{ij}\,,\quad  j_s^{ i}=s v_n^i+\tilde j_s^i\,.
\end{split}
\end{equation}
The argument proceeds as before, including $v_s^i$ in the set of the $O_A$, $X\delta^{ji}$ in the set of the $j^{Ai}$ and $h^i$ in the $s^A$. Then, $s_A j^{Ai}=j_\epsilon^{i}+p v^i+(\mu+v_n\cdot v_s) h^i$ and the ideal components of the currents satisfy
\begin{align}
\label{idealorderU1}
j_{\text{hs}}^{Ai}\partial_i s^A=\partial_i\left[pv_n^i+ (\mu+v_n\cdot v_s) h^i\right]
\end{align}
while
\begin{equation}
\label{DeltaSuperfluids}
\frac{\Delta}{T}=\frac{\Delta_{0}}{T}-\tilde X\partial_i h^i\,
\end{equation}
where $\Delta_0$ is the form of entropy production in the absence of superfluidity, i.e. \eqref{DeltaChargedFluids}. The derivative corrections of superfluids in the absence of boost symmetry have not been worked out in general. There is in principle a class of first-order transport coefficients in the entropy current besides the canonical ones, made of gradients of antisymmetric tensors, such as $h^{[i} v_n^{j]}$. However, such terms have vanishing divergence and so do not contribute either to the quadratic form $\Delta$ or to the matrix $M$.

\subsection{Crystalline solids}

In the case of spontaneously broken translations in all spatial directions, \cite{chaikinlubensky1995,Baggioli:2022pyb} (see also \cite{Delacretaz:2017zxd,Armas:2019sbe,Armas:2020bmo,Armas:2022vpf}), the free energy includes the displacements $u^i$ along the direction $x^i$ through the non-linear strain $u_{ij}=\partial_{(i}u_{j)}+(1/2)\partial_i u^k\partial_j u^k$. For simplicity we work in $d=2$ and so there are two independent displacements. The first law is
\begin{equation}
\label{1stlawTrans}
d\epsilon=Tds+\mu dn+v_{i}dg^i+h_{ij} d(\partial^i u^j)\,,
\end{equation}
where $h_{ij}\equiv\delta f/\delta(\partial_i u_j)$. $f=(1/2)B(u_i^i)^2+G[u_{ij}u^{ij}-(1/2)(u_i^i)^2]$ is the free energy density, $B$ and $G$ the bulk and shear moduli, which quantify the cost of phonon fluctuations in the free energy. For simplicity, we do not consider background strain (the analogue of nonzero background superfluid velocities), which would lead to a more complicated free energy  (see \cite{Armas:2019sbe, Armas:2020bmo, Baggioli:2022pyb}).
The equations of motion are supplemented by the Josephson relations
\begin{equation}
\partial_t u^i+X^i=0\,,\quad X^i=v^i+v^j\partial_j u^i+\tilde X^i
\end{equation}
or
\begin{align}
\partial_t (\partial^j u^k)+\partial_i(\hat{X}^{ijk})=0, \quad \hat{X}^{ijk} = \delta^{ij}X^k
\end{align}
together with the constitutive relations
\begin{align}
\begin{split}
&j_n^{i}=nv^i+\tilde j_n^i\,,\quad j_s^{i}=s v^i+\tilde j_s^{i}\,,\\
& \tau^{ij}=p\delta^{ij}+h^{ij}+h^{il}\partial^j u_l+v^i g^j+\tilde \tau^{ij}\,, 
\end{split}
\end{align}
We now include $\partial_i u_j$ in the $O_A$, $\hat X^{ijk}$ in the $j^{Ai}$ and $h_{ij}$ in the $s^A$.
The argument proceeds as before, with the modifications $s_A j^{Ai}=j_\epsilon^{i}+p v^i+(v_j+v^l\partial_l u_j) h^{ij}$ and then
\begin{align}
\label{idealorderCrystals}
j^{Ai}_{\text{hs}}\partial_i s^A=\partial_i\left[pv_n^i +(v_j+v^l\partial_l u_j) h^{ij}\right].
\end{align}
We also note that at first order in derivatives and with time-reversal symmetry, the entropy current is purely canonical without extra contributions, \cite{Armas:2019sbe, Armas:2021vku, Armas:2020bmo, Armas:2022vpf}. The quadratic form $\Delta$ reads
\begin{equation}
\label{DeltaCrystals}
\frac{\Delta}{T}=\frac{\Delta_{0}}{T}-\tilde X^j\partial^i h_{ij}\,.
\end{equation}

\section{Linear stability with approximate symmetries}

We now illustrate some examples with $f^A \neq 0$, corresponding to explicit breaking of symmetries. In the absence of spontaneous breaking, explicit breaking of a global symmetry $f^\alpha \neq 0$ causes to leading order the corresponding densities to relax, and to subleading order sources a new set of transport coefficients which generally contribute to the real part of modes rather than to their imaginary part, as do more familiar derivative corrections like viscosities and diffusivities. When parity is broken, e.g. by magnetic field, the same logic applies, although the leading order effect is non-dissipative: cyclotron motion.

\subsection{Weak breaking of translations}

As a concrete example, we first consider the weak breaking of translations. This applies in clean metals, through e.g. impurities or umklapp scattering of electrons off the lattice. Ordinarily, this occurs on timescales comparable or shorter than electron-electron scattering, so that the low-energy effective theory needs to keep track of an infinite set of emergent conservation laws for densities on the Fermi surface. In strongly-correlated metals, electron-electron scattering may occur on much shorter timescales than electronic momentum is relaxed. This regime is now accessible in very clean Graphene samples, and may also be relevant in other strongly-correlated metals \cite{Lucas:2017idv,fritz2023hydrodynamic}. In this case, the momentum conservation and entropy production equations become
\begin{equation}
\label{momrelaxequation}
\partial_t g^i+\partial_j\tau^{ji}=f_{g^i}\quad \Rightarrow\quad \frac{\Delta}{T}=\frac{\Delta_{0}}{T}-f_{g^i} \delta v^i
\end{equation} 
Only the deviation out-of-equilibrium $\delta v^i\equiv v^i-\bar v^i$ appears in the divergence of the entropy current. We make this more precise in the Appendix by explicitly considering an external source $\Phi(x)$, which contributes as $-O_\Phi d\Phi/dt$ in \eqref{momrelaxequation}. Integrating over $\Phi$ effectively leaves a $\delta v^i f_{g^i}$ term. Hydrostatic corrections in the absence of external fields capture deviations of the densities $n$, $s$, $\epsilon$, $g^i$ from homogeneous thermal equilibrium and can be absorbed into a redefinition of these quantities and of the first law. Once this is done, they do not explicitly contribute to the equations of motion or cause entropy production. Thus, only out-of-equilibrium corrections need to be added to the constitutive relations.
This implies $\delta v^i$  should now be added to the vector of non-hydrostatic sources $\partial^i s^A$ and $f_{g^i}$ should be expanded order by order in $\{\partial^i s^A,\delta v^i\}$, same as the non-ideal spatial fluxes $\tilde j^{Ai}$. 

To wit, omitting for simplicity nonlinear terms in the fluid velocity, $g^i=\rho v^i+\ell_{eb}\tilde\lambda_v \delta v^i+\mathcal O(\partial,\ell_{eb}^2)$, $f_{g^i}=-\ell_{eb}\rho\tilde\Gamma \delta v^i-\ell_{eb}\tilde\lambda_n\partial^i\mu-\ell_{eb}\tilde\lambda_s\partial^i T+\mathcal O(\partial^2,\ell_{eb}^2)$, $j_n^i=nv^i+\ell_{eb}\tilde\lambda_n \delta v^i+\mathcal O(\partial,\ell_{eb}^2)$, $j_s^i=sv^i+\ell_{eb}\tilde\lambda_s \delta v^i+\mathcal O(\partial,\ell_{eb}^2)$. A gapped mode appears in the spectrum $\omega=-i\ell_{eb}\tilde\Gamma+O(q^2,\ell_{eb}^2)$, allowing to identify $\tilde\Gamma$ as the leading contribution to the momentum relaxation rate. For momentum to relax parametrically slower than other (microscopic) gapped modes, we take $\ell_{eb}\sim\omega\sim q\ll T$. 

The $\tilde\lambda_{s,n,v}$ terms transform under time-reversal identically to the current in which they appear, and so do not cause entropy production.\footnote{Such terms have been discussed in \cite{Armas:2021vku,Armas:2022vpf} and indeed are necessary to match similar contributions found in previous analyses of momentum-relaxed holographic systems \cite{Davison:2015bea,Blake:2015epa,Blake:2015hxa}, as elaborated upon in \cite{Gouteraux:2023uff}. In fact, in the present work we are essentially showing that these terms are present generally without the need to make any specific assumption on the mechanism of momentum relaxation.} All the tilded coefficients contribute to the matrix $\tilde Q$ in \eqref{Mform}, with $\tilde \Gamma$ appearing in $\tilde Q^{(AB)}$ and $\tilde \lambda_{s,n,v}$ in $\tilde Q^{[AB]}$, implying $\tilde\Gamma\geq0$ and no constraints on $\tilde\lambda_{s,n,v}$. Hydrostatic terms contribute to $Q$. 

If parity is not preserved, for instance if a magnetic field is turned on and focusing on 2+1 dimensions for simplicity, then Hall-like terms are generated in the constitutive relations besides the terms we wrote above: $f_{g^i}\supset \ell_{eb}\Gamma_H \epsilon^{ij}v_j+\ell_{eb}\tilde\lambda_n^H\epsilon^{ij}\partial_j\mu+\ell_{eb}\tilde\lambda_s^H\epsilon^{ij}\partial_jT$, etc., which fit into the various matrices we defined in our general discussion. In the special case of a magnetic field, $f_{g^i}=F^{ik}j_k$ is given exactly by the Lorentz force, with $F^{ij}=B\epsilon^{ij}$ and where $B\sim\ell_{eb} \sim \partial$ is the external magnetic field and $\epsilon^{ij}$ the antisymmetric tensor with $\epsilon^{xy}=1$. The Lorentz term contributes in two ways. First, as $Q_0^{[AB]}=TBn\epsilon^{ij}\delta_{g^i}^{[A}\delta_{g^j}^{B]}$ to equation \eqref{Mform}, fixing $\Gamma_H=n B$. Second, it shifts $\partial_i\delta(\mu/T)\mapsto\partial_i\delta(\mu/T)- \delta (v^j/T) F_{ij}$ in \eqref{DeltaChargedFluids}, which contributes in a fixed way to $\tilde\Gamma$ and the $\tilde\lambda_{n,s}$, $\tilde\lambda_{n,s}^H$.
Following our general discussion, this leads to cyclotron modes at $\omega = \pm \omega_c-i\Gamma_c+\mathcal{O}(q)$ but does not affect stability, as long as the matrix $\tilde N$ is positive-definite. 

\subsection{Pinned superfluids}

Explicit breaking of a symmetry in a phase where the symmetry is also spontaneously broken presents qualitatively different features, chiefly the pinning of the Goldstone mode. For simplicity, we will focus here on the case of pseudo-spontaneous breaking of a U(1) symmetry, \cite{Delacretaz:2021qqu,Armas:2021vku,Ammon:2021pyz,Armas:2023tyx,Baggioli:2023tlc}, but our discussion is more general and also applies to e.g. pinned charge density waves and Wigner crystals \cite{Delacretaz:2016ivq,Delacretaz:2017zxd,Delacretaz:2019wzh,Delacretaz:2021qqu,Armas:2021vku,Baggioli:2022pyb,Armas:2023tyx}.

We start by considering a fluid with explicit breaking of a U(1) symmetry. Then, the continuity equation is
\begin{equation}
\label{densityrelaxequation}
\partial_t n+\partial_i j^i=f_{n}\quad \Rightarrow\quad \frac{\Delta}{T}=\frac{\Delta_{0}}{T}-f_{n}\delta\mu
\end{equation} 
where we are directly parametrizing the extra term in the divergence of the entropy current as the out-of-equilibrium contribution to the chemical potential. We can justify this more carefully along the lines of the discussion in \cite{Armas:2021vku} by explicitly including an operator $\Phi$ charged under the U(1), in which case this term would read $-O_\Phi(\mu+d\Phi/dt)$. Integrating out $\Phi$ leaves the $\delta\mu f_n$ term in \eqref{densityrelaxequation}. The analysis proceeds as in the case of relaxed momentum with appropriate modifications.

Since the U(1) is explicitly broken by a charged operator, then in equilibrium the Goldstone field tends to align itself with the background value of the phase of $\Phi$: $\bar\varphi=\psi$, \cite{Armas:2021vku}. Thus new out-of-equilibrium terms proportional to $\delta\varphi=\varphi-\bar\varphi$ are allowed. The leading order term appears as a mass correction to the pressure, $p\mapsto p-(1/2) n_s m^2 \delta\varphi^2$, and leads to extra terms in the first law $\delta p\mapsto \delta p -\delta(n_s m^2 \delta\varphi^2)/2$. It is similar in spirit to the $\tilde\lambda_v$ term in $g^i$ discussed above: this is an out-of-equilibrium term which effectively appears to correct thermodynamics, although most of these corrections are nonlinear in $\delta\varphi$. In \eqref{DeltaSuperfluids}, $\partial_i h^i\mapsto \partial_i h^i-n_s m^2\delta\varphi$.

Our previous arguments for linear dynamical stability then proceed as before.

\subsection{Pinned crystals}

The emergent translation symmetry of electronic charge density waves or Wigner crystals can be broken explicitly by disorder. In this case, the Goldstones of spontaneously broken translations are gapped. If disorder is weak, then the size of the gap is small and they remain light degrees of freedom, which can be retained in the effective field theory description. Here we follow the notation of \cite{Delacretaz:2021qqu,Baggioli:2022pyb} (see also \cite{Delacretaz:2016ivq,Delacretaz:2017zxd,Delacretaz:2019wzh,Armas:2021vku}). Since translations are weakly broken, the shift symmetry of the Goldstone fields is only approximate. In this case, it is better to work with the displacements $u^i$ rather than the strains. Indeed, the free energy density receives an extra contribution $\delta f=(1/2) G q_o^2 \delta u_i \delta u^i$, where $q_o$ is the mass of the pseudo-Goldstones and $\delta u^i$ the out-of-equilibrium deviation of the displacement. The momentum conservation equation is still given by \eqref{momrelaxequation}, while the first law becomes
\begin{equation}
\label{firstlawpinned}
d\epsilon=Tds+\mu dn+v_{i}dg^i+h_{ij} d(\partial^i u^j)+G q_o^2 \delta u_i d\delta u^i\,.
\end{equation}
The next steps proceed similarly as in the spontaneous case, with $s_A j^{Ai}$ and  \eqref{idealorderCrystals} still the same. We find extra contributions to $\Delta$ compared to the case of momentum-relaxed fluids \eqref{momrelaxequation}
\begin{equation}
\label{DeltaPinned}
\frac{\Delta}{T}=\frac{\Delta_{0}}{T}+v_i \tilde f_{g^i}-\tilde X^j\left(\partial^i h_{ij}-G q_o^2 \delta u_j\right)\,.
\end{equation} 
This leads to a change in the dissipative sources, $\partial_is^A\mapsto\partial_is^A-G q_o^2\delta  u^i\delta^A_{u^i}$. In other words, pinning modifies the dissipative combination multiplying the matrix $\tilde{N}$ in the constitutive relations at first order in gradients. $\tilde{N}$ itself must also be enlarged as for relaxed fluids \eqref{momrelaxequation} to include a constitutive relation for $\tilde f_{g^i}\equiv f_{g^i}-G q_0^2 (\delta u^i+\delta u^j\partial^i\delta  u_j)$ while the set of non-hydrostatic sources is enlarged to include $\delta v^i$.

\section{Acknowledgments}

\begin{acknowledgments}

We thank Filippo Sottovia for collaboration at an early stage of this work.
We are grateful to Ashish Shukla, Vaios Ziogas and Benjamin Withers for discussions. We thank Richard Davison for comments on an earlier version of this manuscript. The work of B.~G.~ was supported by the European Research Council (ERC) under the European Union's Horizon 2020 research and innovation programme (grant agreement No758759). The work of E.~M.~was supported in part by NSERC and in part by the European Research Council (ERC) under the European Union's Horizon 2020 research and innovation programme (grant agreement No758759). We also acknowledge the hospitality of the Kavli Institute for Theoretical Physics where part of this work was performed and which is supported in part by the National Science Foundation under Grants No.~NSF PHY-1748958 and PHY-2309135.

\end{acknowledgments}

\bibliography{biblio}

 \newpage
 
\appendix

\section{Modern formulation of hydrodynamics and the off-shell second law}

In this section, we provide explicit derivations for many of the relations in the main text which follow from the construction of an effective action constrained by diffeomorphism and gauge symmetries. Such an approach has been called the `modern form' of hydrodynamics.

In the modern formulation of hydrodynamics \cite{Loganayagam:2011mu, Haehl:2015pja}, hydrodynamics is split up into two steps: 1) determining the hydrostatic constitutive relations via a generating functional in terms of background sources and 2) determining out-of-equilibrium constitutive relations via an off-shell second law written in terms of a free energy current, $N^\mu$. Our presentation in the main text summarizes both of these steps, in the limit that the background fluid manifold (or spacetime) is flat, i.e. its metric is $\eta_{\mu\nu} = \text{diag}(-1,1,1,1...)$ for $x^\mu = \{t,x^i\}^\mu$ and $i = 1,...,d$. We will also occasionally use $v^\mu = \{1,v^i\}^\mu$. For our purposes, we do not couple the fluid to curved spacetime, since this is only necessary to find correlation functions of momentum and the stress tensor and not to determine the linearized spectrum of modes.\footnote{This means that our results also apply to boost-agnostic fluids which couple to Aristotelian spacetimes, since these too have a flat limit in which the metric is $\eta^{\mu\nu}$.} 

Hydrodynamics is a low energy effective field theory that can be described by an effective action $S$ that couples conserved currents, $j^{\alpha\mu}$ to background fields $B_{\alpha\mu}$, and can include Goldstone fields\footnote{We assume that only the `superfluid velocities' appear in the action, but include a $\phi^{at} = 0$.} $\phi^{a\mu}$ and explicit symmetry breaking fields $\psi^I$ that do not directly couple to conserved currents. The index $\alpha = 1,...,L$ covers the $L$ conserved charges, $a=1,...,M$ covers the $M$ Goldstones, and the index $I=1,...,N$ covers the $N$ symmetry breaking fields with a combined index $A=1,...,L+M+N$. The effective action is
\begin{align}
\label{eq:effective_action_variation}
\begin{split}
\delta S &= \int d^{d+1}x \, \biggl(j^{\alpha \mu}\delta B_{\alpha \mu} +X^{a \mu}\delta \phi_{a \mu}+ \Psi^I \delta \psi_{I} \biggr)\\
&= \int d^{d+1}x\, J^{A\mu} \delta S_{A\mu}
\end{split}
\end{align}
Here, we have included conserved densities as $j^{\alpha t} = O^\alpha$. We have also defined the generalized currents $J^{A\mu}$ to span the current $j^{\alpha\mu}$, $X^{a\mu}$, and $\Psi^I$ with generalized sources $S_{A\mu}$ spanning $B_{\alpha\mu}$, $\phi_{a\mu}$ and $\psi_I$. 

Since the $\phi_{ a\mu}$ are dynamical, they couple to their equations of motion
\begin{align}
X^{a \mu} = 0 \,.
\end{align}
If there is no spontaneous symmetry breaking, we can impose this as an identity and set $\phi_{a \mu}=0$. Global symmetries of the fluid transform $B_{\alpha \mu}, \phi_{a\mu}$ and $\psi_I$. These symmetries include diffeomorphisms, defined by a vector $\beta^\mu$, and potentially other global symmetries if there are conserved charges. For the sake of simplicity, we will ignore non-abelian global symmetries and include just an abelian parameter $\lambda_{A}$. We expect that our results hold equally well for non-abelian symmetries, though the transformation laws are more involved. Working in flat spacetime, the transformations of the background fields read
\begin{align}
\label{eq:global_symmetries}
\delta_{\mathcal{B}} B_{\alpha \mu} = \pounds_\beta B_{\alpha \mu} + \partial_\mu \lambda_\alpha
\end{align}
Among the background fields is the metric leading to the transformation properties
\begin{align}
\begin{split}
\delta_{\mathcal{B}} B_{\epsilon\mu} &= \eta_{t\nu}\partial_\mu \beta^\nu = - \partial_\mu \beta^t, \, , \\
\delta_{\mathcal{B}} B_{g_i\mu} &= \eta_{i\nu}\partial_\mu \beta^\nu =  \partial_\mu \beta^i\,.
\end{split}
\end{align}

When one or more of the global symmetries is spontaneously broken, we have
\begin{align}
\delta_{\mathcal{B}} \phi_{a \mu} = \pounds_\beta \phi_{a \mu} - g^a_{\alpha}\partial_\mu \lambda_{\alpha}
\end{align}
where $g^a_{\alpha}$ is a constant so that the gauge invariant combination $\xi_{a\mu} = \phi_{a\mu}+g^a_{\alpha}B_{\alpha\mu}$ satisfies
\begin{align}
\delta_{\mathcal{B}} \xi_{a\mu} = \pounds_\beta \xi_{a\mu} = \partial_\mu (\beta^\nu\xi_{a\nu}) +g^a_{\alpha} \beta^\nu F^\alpha_{\nu\mu}
\end{align}
where $F^{\alpha}_{\mu\nu} = 2\partial_{[\mu}B_{\nu]}^\alpha$ and we use that $\partial_{[\mu}\phi_{\nu]}^a = 0$. Since $F^\epsilon = F^{g_i} = \lambda_\epsilon = \lambda_{g_i} = 0$, we can write 
\begin{align}
\delta_{\mathcal{B}} B_{\alpha \mu} = \beta^\nu F^{\alpha}_{\nu\mu} + \partial_\mu(\lambda_{\alpha}+\beta^\nu B_{\alpha\nu}) 
\end{align}

If there are symmetries such that $B_\alpha$ includes more spacetime indices, then we must be more careful, but our results would nevertheless be the same. An example would be spontaneously broken spacetime symmetries, like in a crystal. We cover this in the main text, so ignore a general treatment here. Finally, if there are symmetry breaking fields $\psi_I$, they must also be invariant under diffeomorphisms leading to
\begin{align}
\delta_{\mathcal{B}}  \psi_I = \beta^\mu\partial_\mu \psi_I.
\end{align}

We define the `thermodynamic frame' as a relation between the symmetry parameters and the sources defined in the main text as
\begin{align}
\label{eq:thermodynamic_frame}
\begin{split}
s_{\alpha} = \lambda_{\alpha } + \beta^\mu B_{\alpha \mu} , \quad \beta^t = \frac{1}{T}, \quad \beta^i = \frac{v^i}{T}\,.\end{split}
\end{align}
This includes $\alpha = \epsilon$ with $s_\epsilon = -\beta^t$ and $\alpha = g_i$ with $s_{g_i} = \beta^i$. Note that with these definitions,
\begin{align}
\begin{split}
\delta_{\mathcal{B}} B_{\alpha \mu} &= \beta^\nu F_{\nu\mu}^\alpha  + \partial_\mu s_{\alpha } \,,\\
\delta_{\mathcal{B}}\phi_{a\mu} &= \partial_\mu(\beta^\nu\xi_{a\nu}-g^a_{\alpha}s^\alpha) = \partial_\mu (\delta_{\mathcal{B}}\phi_a)\,\\
&= \delta_\mathcal{B}\xi_{a\mu} - g^a_{\alpha}\delta_{\mathcal{B}}B_{\alpha\mu} 
\end{split}
\end{align}
where in the last equality we used that $\phi_{a\mu} = \partial_\mu\phi_a$.

These symmetries must be symmetries of $S$, and inserting \eqref{eq:global_symmetries} into \eqref{eq:effective_action_variation}  we have
\begin{align}
\begin{split}
\delta S &= \int d^{d+1}x\, \biggl(-s_{\hat{\alpha}}[\partial_\mu j^{\hat{\alpha}\mu}-g^{\hat{\alpha}a}X^a] \\
&- s_{\epsilon}[\partial_\mu j^{\epsilon \mu}+F^{\alpha}_{t\mu}j^{\alpha\mu}-\xi_{a t}X^a+\Psi^I\partial_t\psi^I]\\
&-s_{g^i}[\partial_\mu j^{g^i\mu} - F^{\alpha}_{i\mu}j^{\alpha\mu}+\xi_{a i}X^a-\Psi^I\partial_i\psi^I]\\
&+\text{total derivative}\biggr) \,,
\end{split}
\end{align}
where we defined $X^a= \partial_\mu X^{a\mu}$ for ease of reading and $\hat{\alpha}$ spans all $\alpha$ indices except $\epsilon$ and $g_i$. We can ignore the total derivative and $\delta S = 0$ imposes that the terms in square brackets vanish identically, which give the equations of motion \eqref{eoms} with the identification
\begin{align}
\begin{split}
f^{\hat{\alpha}}&=g^{\hat{\alpha}a}h^a \, , \\
f^{\epsilon} &= -F_{t\mu}^{\alpha }j^{\alpha \mu} + \xi_{a t}X^a  - \Psi^I\partial_t \psi^I \, , \\
f^{g^i} &= F_{i\mu}^{\alpha }j^{\alpha \mu}  - \xi_{a i}X^a+\Psi^I\partial_i \psi^I  \,.
\end{split}
\end{align}

The hydrostatic constitutive relations for the densities and fluxes can be obtained via a generating functional $\mathcal{F}$ which is a gauge-invariant scalar function of the background and hydrodynamic fields. In particular, for symmetry transformations,
\begin{align}
\label{eq:delta_B_F_1}
\begin{split}
\delta_{\mathcal{B}} \mathcal{F} &= \left(\frac{\delta \mathcal{F}}{\delta B_{\alpha \mu}}\right)\delta_{\mathcal{B}} B_{\alpha \mu} +\left(\frac{\delta \mathcal{F}}{\delta \phi_{a\mu}}\right)\delta_{\mathcal{B}} \phi_{a\mu} \\
&\quad+ \left(\frac{\delta \mathcal{F}}{\delta \psi_{I}}\right)\delta_{\mathcal{B}} \psi_I+  \partial_\mu\Theta_\mathcal{F}^\mu 
\end{split}
\end{align} 
Here, the terms in parentheses are the Euler-Lagrange derivatives of $\mathcal{F}$ which can lead to a total derivative $\Theta_\mathcal{F}$ due to integration by parts.
Then, using $\delta_{\mathcal{B}} \mathcal{F} = \beta^\mu \partial_\mu \mathcal{F}$, 
\begin{align}
\begin{split}
\label{eq:generating_functional}
-\partial_\mu (\mathcal{F}\beta^\mu) &= j^{\alpha \mu}_{\text{hs}}\delta_{\mathcal{B}} B_{\alpha \mu} +X^{a\mu}_{\text{hs}}\delta_{\mathcal{B}} \phi_{a\mu} + \Psi^I \delta_{\mathcal{B}} \psi_I
 -\partial_\mu \Theta_{\mathcal{F}}^\mu
\end{split}
\end{align}
with
\begin{align}
\begin{split}
\label{eq:hydrostatic_const_rel}
j^{\hat{\alpha }\mu}_{\text{hs}} &= -\frac{\delta \mathcal{F}}{\delta B_{\hat{\alpha }\mu}} \, ,  \quad X^{a\mu}_{\text{hs}} = -\frac{\delta \mathcal{F}}{\delta \phi_{a\mu}} \,, \quad  \Psi_I = - \frac{\delta\mathcal{F}}{\delta \psi_I} \, ,\\ 
j^{\epsilon\mu}_{\text{hs}} &=-\frac{\delta \mathcal{F}}{\delta B_{\epsilon\mu}}+\mathcal{F}\delta^{t\mu} \, , \quad
j^{g_i\mu}_{\text{hs}} = -\frac{\delta \mathcal{F}}{\delta B_{g_i\mu}}-\mathcal{F}\delta^{i\mu} \,.
\end{split}
\end{align}

The hydrostatic constitutive relations in \eqref{eq:hydrostatic_const_rel} must be consistent with the absence of entropy production. To arrive at this conclusion, the second law
\begin{align}
\label{appsecondlaw}
\partial_t s + \partial_i j_s^i = \Delta/T
\end{align}
is first reformulated to apply for off-shell configurations of the sources by adding linear combinations of the equations of motion,
 \begin{align}
 \label{appsecondlawoffshell}
 \partial_\mu j_s^\mu + \mathcal{S}_\alpha  (\partial_\mu j^{\alpha \mu} - f^{\alpha })  = \Delta/T \geq 0 \,
 \end{align}
where $\mathcal{S}_\alpha $ are termed `Lagrange multipliers' and can be adopted as out-of-equilibrium definitions for the sources. The indices $A$ span the indices $\hat{A}, \epsilon, g_i$ and $\alpha$. On-shell, where the equations of motion are satisfied, it is clear that this returns \eqref{appsecondlaw}. Importantly, we have not included the explicit external symmetry breaking terms $\psi_a$ except through their contribution in $f^\alpha$. 

Now, we define
 \begin{align}
 N^\mu \equiv j_s^\mu + \mathcal{S}^\alpha j^{\alpha \mu} +\Xi^a X^{a\mu}+ \Theta^\mu
 \end{align}
 where $\Theta^\mu$ is a term which can arise via integration by parts in the second law. This is observed to satisfy
 \begin{align}
 \partial_\mu (N^\mu-\Theta^\mu-\Xi^a X^{a\mu}) = \Delta/T +\mathcal{S}^\alpha f^\alpha  + j^{\alpha \mu}\partial_\mu \mathcal{S}^\alpha 
 \end{align}
 which is strikingly similar to \eqref{Deltaeq}. In fact, if we take $\mathcal{S}^\alpha = s^\alpha $ and $\Xi^a = 0$ we have
 \begin{align}
 \label{eq:2nd-law-thermodynamic-frame1}
 \begin{split}
& \partial_\mu(N^\mu - \Theta^\mu-\Xi^a X^{a\mu})\\
 & = \Delta/T +j^{\alpha\mu}\delta_{\mathcal{B}}B_{\alpha\mu} - X^{a}\delta_{\mathcal{B}}\phi_a + \Psi^I\delta_{\mathcal{B}}\psi_{I}\,.
 \end{split}
 \end{align}
 If instead we choose $\Xi^a = \delta_{\mathcal{B}}\phi_a$, then using $\partial_\mu X^{a\mu} = X^a$,
 \begin{align}
  \label{eq:2nd-law-thermodynamic-frame2}
 \begin{split}
 \partial_\mu (N^\mu-\Theta^\mu) &= \Delta/T + j^{\alpha\mu}\delta_{\mathcal{B}} B_{\alpha\mu}+X^{a\mu}\delta_{\mathcal{B}} \phi_{a\mu} \\
 &\quad+\Psi^I\delta_{\mathcal{B}} \psi_I\, ,\\
 &= \Delta/T + J^{A\mu}\delta_{\mathcal{B}}S_{A\mu} \,.
 \end{split}
 \end{align}
 Then, by inspection, choosing
 \begin{align}
 \label{eq:hydrostatic_freeenergy}
 \begin{split}
 N^\mu_{\text{hs}} &= -\mathcal{F}\beta^\mu + \Theta^\mu_{\mathcal{F}}\,, \quad \Theta^\mu_{\text{hs}} = \Theta^\mu_{\mathcal{F}} \,,\\
  j_{s,\text{hs}}^\mu &= -\mathcal{F}\beta^\mu - s_\alpha j^{\alpha \mu}_{\text{hs}} - X^{a\mu}_{\text{hs}}\delta_{\mathcal{B}}\phi_a
  \end{split}
 \end{align}
 together with \eqref{eq:hydrostatic_const_rel} implies $\Delta = 0$,  so that the hydrostatic constitutive relations do not lead to entropy production. Note that either \eqref{eq:2nd-law-thermodynamic-frame1} or \eqref{eq:2nd-law-thermodynamic-frame2} are equally valid and differ only in the choice of $N^\mu_{\text{hs}}$ and therefore $j_{s,\text{hs}}^{\mu}$. 

In equilibrium, there exist preferred symmetry parameters $\beta^\mu \to \bar{\beta}^\mu = \bar{v}^\mu/\bar{T}$ and $\lambda_{\alpha} \to \bar{\lambda}_{\alpha} = \bar{\mu}_{\alpha}/\bar{T} - \bar{v}^\mu B_{\alpha\mu}/\bar{T}$, such that 
\begin{align}
\delta_{\mathcal{K}}\mathcal{F} = \delta_{\mathcal{K}} B_{\alpha \mu} = \delta_{\mathcal{K}} \phi_{a\mu} = \delta_{\mathcal{K}} \psi_{I} = 0.
\end{align}
Thus in hydrostatic equilibrium, a stricter condition holds, with $\Delta = 0$ and $\partial_\mu (-\mathcal{F} K^\mu) = 0$. Nevertheless, the hydrostatic constitutive relations extend out-of-equilibrium when they are derived from the hydrostatic partition function with parameters $\beta^\mu$ and $\lambda_{\alpha}$. However, the existence of thermal equilibrium is important for constraining the form of the hydrostatic partition function $\mathcal{F}$ which is required to be constructed from gauge invariant scalars which are not constrained to vanish in equilibrium. In particular, any scalar explicitly constructed from $\delta_{\mathcal{B}}(\cdot)$ will necessarily vanish. One consequence is that because $\delta_{\mathcal{B}}\phi_a = \beta^\mu\xi_{a\mu}-g^a_\alpha s^{\alpha}$, some linear combinations of $\beta^t\xi_{at}$, $\beta^i\xi_{a i}$, and $s_{\alpha}$ will vanish. Hence, we can choose a linearly independent set which excludes $\beta^t\xi_{at}$. As a consequence, $X_{\text{hs}}^{at} = 0$ and $\xi_{at}$ will not appear (or can be eliminated) in the hydrostatic constitutive relations. Furthermore, one can always eliminate $\partial_t s_{\alpha}$ and $F^\alpha_{ti}$ in favor of $F^\alpha_{ij}$ and $\partial_i s_\alpha$. As an example, when this is taken into account, in the absence of background fields and identifying $\mathcal{F}=-p$, $O^\alpha=j^{\alpha t}_{hs}$ and $X^{ai}_{\text{hs}} = h^{ai}_{\text{hs}}/T$,
\begin{align}
\begin{split}
s+\mathcal{O}^\alpha s_{\alpha} &= p/T \,,\\
j^i_{s,\text{hs}} + j^{\alpha i}s_{\alpha}&= \frac{p}{T}v^i - (Tg^a_{\alpha}s^{\alpha} - v^i\partial_i\phi_a)\frac{h^{ai}_{\text{hs}}}{T}
\end{split}
\end{align}
matching the main text. 

The hydrostatic constitutive relations naturally lead to a first law of thermodynamics. Instead of background fields, we could equally well have used the gauge invariant sources, exchanging $B_{\alpha \mu}$ for $s_{\alpha}$ and $\xi_{a \mu}$. Adopting $\delta_{\mathcal{B}}\phi_{a,hs} = 0$, we have $\delta_{\mathcal{B}}\xi_{a\mu} = (g^{-1})^{\alpha}_a\delta_{\mathcal{B}}B_{a\mu}$. Then, we can write
\begin{align}
\begin{split}
\delta_{\mathcal{B}}\mathcal{F} &= (\frac{\delta \mathcal{F}}{\delta B_{\alpha\mu}}+g^\alpha_a X^{a\mu})\delta_{\mathcal{B}}B_{\alpha\mu} - X^{a \mu}\delta_{\mathcal{B}}\xi_{a\mu} \\
&+ \Psi^I\delta_{\mathcal{B}}\psi_I + \partial_\mu\Theta^\mu_{\mathcal{F}} \,.
\end{split}
\end{align}
and for this to be gauge invariant, we note that $\beta^\mu\delta_{\mathcal{B}}B_{\alpha\mu} = \delta_{\mathcal{B}}s_\alpha$ requires
\begin{align}
\label{jmu_hydro_def}
\frac{\delta\mathcal{F}}{\delta B_{\alpha\mu}}+g^{\alpha}_aX^{a\mu} = \rho_n^{\alpha}\beta^\mu
\end{align}
for some $\rho_n^{\alpha}$. This leads to,
\begin{align}
\delta_{\mathcal{B}}\mathcal{F} = \rho_n^\alpha\delta_{\mathcal{B}}s_{\alpha} - X^{a \mu}\delta_{\mathcal{B}}\xi_{a\mu} + \Psi^I\delta_{\mathcal{B}}\psi_I + \partial_\mu\Theta^\mu_{\mathcal{F}} \,.
\end{align}
We also want everything on the right hand side to be scalars under diffeomorphisms (up to the total derivative) which requires
\begin{align}
\label{Xmu_hydro_def}
X^{a\mu}_{\text{hs}} = -2\rho_s^{(ab)}\xi^{b\mu}
\end{align}
so that we have
\begin{align}
\delta_{\mathcal{B}}\mathcal{F} = \rho_n^\alpha\delta_{\mathcal{B}}s_{\alpha} + \rho_s^{(ab)}\delta_{\mathcal{B}}(\xi_{a}\cdot\xi_{a}) + \Psi^I\delta_{\mathcal{B}}\psi_I + \partial_\mu\Theta^\mu_{\mathcal{F}} \,.
\end{align}
Since $\beta^\mu\xi_{a\mu} = g^a_\alpha s^{\alpha}$, it is useful to split up
\begin{align}
\xi_a\cdot\xi_b = (v^i \xi_{ai}-g^{a}_{\alpha}s^{\alpha})(v^j \xi_{bj}-g^{b}_{\beta}s^{\beta}) + (\xi_{ai}\xi^{bi})
\end{align}
Leading to 
\begin{align}
\delta_{\mathcal{B}}\mathcal{F}= \rho^{\alpha}\delta_{\mathcal{B}}s_{\alpha} + h^{ai}\delta_{\mathcal{B}}\xi_{ai} + \Psi^I\delta_{\mathcal{B}}\psi_I + \partial_\mu \Theta^\mu_{\mathcal{F}}
\end{align}
where
\begin{align}
\label{rhoandh_defs}
\begin{split}
\rho^\alpha &= \rho_n^{\alpha} + 2\rho^{(ab)}_sg^{a}_{\alpha}g^{b}_{\beta}s^\beta - 2\rho^{(ab)}v^i\xi_{ai}g^{b}_{\alpha}\\
&+2T\rho^{(ab)}_s\xi_{ai}(v^j\xi_{bj}-g^b_\beta s^\beta)(\delta^{\alpha g_i}+v^i\delta^{\alpha\epsilon})\,\\
h^{ai} &= 2\rho^{(ab)}_{s}(\xi^{bi} + v^i[v^j\xi_{bj}-g^{b}_\beta s^\beta])\,.
\end{split}
\end{align}
Finally, using
\begin{align}
O^{\alpha}_{\text{hs}} = j^{\alpha t}_{\text{hs}}
\end{align}
we find
\begin{align}
O_{\text{hs}}^\alpha \equiv  \frac{\delta (-\mathcal{F}\beta^t)}{\delta s^\alpha} \,.
\end{align}
This tells us that $\mathcal{F} = -p$ and we have the first law
\begin{align}
d(p/T) = O^\alpha ds_\alpha - \frac{h^{ai}}{T}d\xi_{ai}-\frac{\Psi^I}{T} d \psi_I -\Theta_d\,.
\end{align}
This matches the main text up to the term $\Theta_d$ which can depend on the type of variation of $p/T$. Since we have demonstrated the first law and the Josephson relations are consistent via the hydrostatic generating functional approach and the standard approach, the currents shown in the main text are consistent as well. These currents are obtained by using \eqref{rhoandh_defs} in \eqref{jmu_hydro_def} and \eqref{Xmu_hydro_def}.

As a simple example, we can consider the case with no spontaneous symmetry breaking, in which case
\begin{align}
\frac{\delta \mathcal{F}}{\delta B_{\alpha \mu}} = \rho^\alpha  \beta^\mu
\end{align}
so that using \eqref{eq:hydrostatic_const_rel}
\begin{align}
\begin{split}
j^{\alpha i}_{\text{hs}} &= (\epsilon+p)v^i\delta^{\alpha\epsilon} + (g_jv^i+p\delta^{ij})\delta^{\alpha g_j}+ O^{\hat{\alpha}}v^i\delta^{\alpha\hat{\alpha}} \,.
\end{split}
\end{align}
where we identified $\rho^\epsilon = -T(\epsilon+p)$, $\rho^{g_i} = -g_i/T$ and $\rho^{\hat{\alpha}} = -O^\alpha/T$. Note that 
\begin{align}
\partial_i j^{\alpha i} \supset \delta^{\alpha \epsilon}v^i\Psi^I\partial_i\psi_I + \delta^{\alpha g_j}\Psi^I\partial_i\psi_I
\end{align}
arising from $\partial_i$ hitting $p$, which cancels the explicit symmetry breaking pieces in the equations of motion coming from scalar fields. Then, we need only consider variations in the equations of motion arising from the hydrodynamic fields.

Now, linearizing around an equilibrium state, $s^{\alpha} =  \bar{s}^{\alpha}+ \delta s^\alpha$, but with $\partial_\mu\bar{s}^\alpha = 0$ (see the next section for relaxing this assumption), we have
\begin{align}
\begin{split}
\partial_i\delta j^{\alpha i}_{\text{hs}} &= N^{(\alpha\beta)i}\partial_i\delta s_{\beta} + f(\psi_I)\,\\
N^{(\alpha\beta)i} &=2\bar{T}\left(p\delta^{\epsilon(\alpha} + \bar{O}^{(\alpha}\right)\left(\delta^{\beta)g_i}+\bar{v}^i\delta^{\beta)\epsilon}\right)
\end{split}
\end{align}
and 
\begin{align}
\delta f^{\alpha} = -\delta\left(\frac{O^C}{T}v^\mu F^C_{[\mu \nu]} \frac{\delta v^\nu}{\delta s^\alpha}\right)+f(\psi_I) \,.
\end{align}
Here, we have written the variations which explicitly depend on $\psi_I$ as $f(\psi_I)$. For simplicity, we will set $\bar{v} = 0$, though relax this constraint in the next section and arrive at the same conclusions. Then,
\begin{align}
\begin{split}
\delta f^{\alpha} &= -Q^{[\alpha\beta]}\delta s_\beta \,, \\ Q^{[\alpha\beta]} &= -\frac{\bar{O}^C}{\bar{T}}\frac{\delta v^\mu}{\delta s^\alpha}\frac{\delta v^\nu}{\delta s^\beta}F^C_{[\mu\nu]} \,.
\end{split}
\end{align}
In other words, 
\begin{align}
\label{linearizedeomhydrostatic}
\partial_i \delta j_{\text{hs}}^{\alpha i} - \delta f^{\alpha} = Q^{[\alpha\beta]}\delta s_{\beta} + N^{(\alpha\beta)i}\partial_i\delta s_{\beta}
\end{align}
When the fluctuations take the form of plane waves, $\partial_i\delta s_\beta = iq_i\delta s_\beta$ and the matrices in \eqref{linearizedeomhydrostatic} are projected into longitudinal components along $q^i$ and can be combined into $i$ times a Hermitian matrix. This remains true, as we show in the main text, when symmetries are spontaneously broken. In fact, though we illustrated results with symmetry breaking scalars, we expect that our results apply to more general symmetry breaking fields--when the symmetry breaking field is not a background gauge field associated to a local symmetry, then its symmetry breaking pieces in $f^A$ can be cancelled in the equations of motion by a contribution to $j^{Ai}_{\text{hs}}$ by suitable modifications of the first law. In other words we may write
\begin{align}
j^{Ai}_{\text{hs}} \supset j^{Ai}_{eb}(\{\psi\}), \quad f^A \supset \partial_ij^{Ai}_{eb}(\{\psi\})\,.
\end{align}
Then, $\delta f^{A}_{\text{hs}}$ can only contain terms which come from background gauge fields which lead to an antisymmetric $Q^{[AB]}$. Out-of-equilibrium, symmetry breaking fields can explicitly appear in the equations of motion, as we now show.

Returning to the off-shell second law, we can separate components into hydrostatic and non-hydrostatic pieces: $j^{\alpha \mu} = j^{\alpha\mu}_{\text{hs}}+j_{\text{nhs}}^{\alpha\mu}$, $h^{a\mu} = X_{\text{hs}}^{a\mu}+X_{\text{nhs}}^{a\mu}$, $\Psi^{I} = \Psi_{\text{hs}}^I+\Psi_{\text{nhs}}^I$, $N^\mu = N^\mu_{\text{hs}}+N^\mu_{\text{nhs}}$, and $\Theta^\mu = \Theta^\mu_{\text{hs}}+\Theta_{\text{nhs}}^\mu$. Since the hydrostatic pieces have already been accounted for, we write
\begin{align}
\begin{split}
\label{nhs_offshellsecondlaw}
\partial_\mu (N_{\text{nhs}}^\mu-\Theta_{\text{nhs}}^\mu) &= \Delta/T +J^{A\mu}_{\text{nhs}}\delta_{\mathcal{B}}S_{A\mu} \,.
\end{split}
\end{align}

The thermodynamic frame relates the parameters $\beta^\mu$ and $\lambda_{\alpha}$ to the sources $s_\alpha$. In equilibrium, these choices are uniquely defined; however, out-of-equilibrium, we may redefine the $\beta^\mu\to (\beta')^\mu$ and $\lambda_{\alpha}\to \lambda_{\alpha}'$, leading to a different definition of out-of-equilibrium sources $s_\alpha'$ (though with the same functional relation between the two, given by \eqref{eq:thermodynamic_frame}). We can hence choose the parameters $\beta^\mu$ and $\lambda_{\alpha}$ such that $j_{\text{nhs}}^{\alpha t} = 0$. Furthermore, we can use a field redefinition of $\phi_a$ such that $X_{\text{nhs}}^{at}=0$. Altogether, these choices exhausts the redefinition freedom, fully fixing the form of $s_\alpha$ out-of-equilibrium and give the `thermodynamic density frame.' Taking this into account, we may write
\begin{align}
\label{eq:second_law_thermodynamic_density_frame}
\begin{split}
\partial_\mu(N_{\text{nhs}}^\mu -\Theta^\mu_{\text{nhs}}) &= \Delta/T + J^{Ai}_{\text{nhs}}\delta_{\mathcal{B}}S_{Ai} \\
&= \Delta/T + j_{\text{nhs}}^{\alpha i}\delta_{\mathcal{B}} B_{\alpha i} \\
&\quad- X_{\text{nhs}}^{a}\delta_{\mathcal{B}} \phi_{a} + \Psi_{\text{nhs}}^I\delta_{\mathcal{B}}\psi_I
\end{split}
\end{align}
or if we treat $\phi_{a\mu}$ as the independent variable, 
\begin{align}
\label{eq:second_law_thermodynamic_density_frame}
\begin{split}
\partial_\mu(N_{\text{nhs}}^\mu -\Theta^\mu_{\text{nhs}}) &= \Delta/T + J^{Ai}_{\text{nhs}}\delta_{\mathcal{B}}S_{Ai} \\
&= \Delta/T + j_{\text{nhs}}^{\alpha i}\delta_{\mathcal{B}} B_{\alpha i} \\
&\quad+ X_{\text{nhs}}^{ai}\delta_{\mathcal{B}} \phi_{ai} + \Psi_{\text{nhs}}^I\delta_{\mathcal{B}}\psi_I \,.
\end{split}
\end{align}

One important set of transport coefficients is fixed by
\begin{align}
X_{\text{nhs}}^{a} = (\sigma_{\phi}^{-1})^{ab}(\delta_{\mathcal{B}}\phi_{b} - \hat{X}^b/T)
\end{align}
where $\sigma_{\phi}^{ab}$ is an operator whose symmetric part under $a\leftrightarrow b$ is positive definite (and hence, invertible) and the $\hat{X}^{b}$ indicate other dissipative contributions. In the $\phi_{a\mu}$ ensemble, we would have
\begin{align}
X_{\text{nhs}}^{ai} = -(\tilde{\sigma}_{\phi}^{-1})^{ab}(\delta_{\mathcal{B}}\phi_{bi} - \hat{X}^{bi}/T)
\end{align}

Since $X^a = 0$ is the Goldstone equation of motion, we have $X^a_{\text{nhs}}=-X^a_{\text{hs}}$. This leads to the `Josephson relation'
\begin{align}
\delta_{\mathcal{B}}\phi_{a} = \sigma_\phi^{ab}X_{\text{hs}}^{b}+\hat{X}^{a}/T\,.
\end{align}
It is natural, then, to treat $X^b_{\text{hs}}$ as a `non-hydrostatic source' (\cite{Arean:2023nnn} does this in practice for relativistic superfluids). In the absence of background sources, the Josephson relation reads
\begin{align}
\begin{split}
&\partial_t\phi_a+v^j\partial_j\phi_{a} - Tg^a_{\alpha}s^{\alpha}+\tilde{X}^a = 0\,,
\end{split}
\end{align}
where we have identified 
\begin{align}
\tilde{X}^a = \hat{X}^a + T\sigma_\phi^{ab}X_{\text{hs}}^b
\end{align}
matching the expression in the main text.

We may write
\begin{align}
\begin{split}
j_{\text{nhs}}^{\alpha i} &= (\mathcal{D}_{OO}+\bar{\mathcal{D}}_{OO})^{\alpha\beta ij}\delta_{\mathcal{B}} B_{\beta j} +(\mathcal{D}_{O\phi} + \bar{\mathcal{D}}_{O\phi})^{\alpha a i}X^a_{\text{hs}}\\
&\quad+(\mathcal{D}_{O\psi}+\bar{\mathcal{D}}_{O\psi})^{aIi}\delta_{\mathcal{B}}\psi_I \,,\\
\delta_{\mathcal{B}}\phi^{a}_{\text{nhs}} &= (\mathcal{D}_{O\phi}-\bar{\mathcal{D}}_{O\phi})^{\alpha a j}\delta_{\mathcal{B}}B_{\alpha j}+(\mathcal{D}_{\phi\phi}+\bar{\mathcal{D}}_{\phi\phi})^{ab}X^b_{\text{hs}} \\
&\quad + (\mathcal{D}_{\phi\psi}+\bar{\mathcal{D}}_{\phi\psi})^{aI}\delta_{\mathcal{B}}\psi_I \, , \\
\Psi^I_{\text{nhs}} &= (\mathcal{D}_{O\psi}-\bar{\mathcal{D}}_{O\psi})^{aIi}\delta_{\mathcal{B}}B_{ai} + (\mathcal{D}_{\phi\psi}-\bar{\mathcal{D}}_{\phi\psi})^{aI}X^a_{\text{hs}}\\
&\quad + (\mathcal{D}_{\psi\psi}+\bar{\mathcal{D}}_{\psi\psi})^{IJ}\delta_{\mathcal{B}}\psi_J
\end{split}
\end{align}
so that all expressions have a similar form, we have defined $\sigma_\phi^{ab} = (\mathcal{D}_{\phi\phi}+\bar{\mathcal{D}}_{\phi\phi})^{ab}$. The $\mathcal{D}$ and $\bar{\mathcal{D}}$ are differential operators. For example,
\begin{align}
\begin{split}
\mathcal{D}_{OO}^{\alpha\beta ij} &= (\mathcal{D}_{OO}^{(0)})^{\alpha\beta ij} + (\mathcal{D}_{OO}^{(1)})^{\alpha\beta ij\mu}\partial_\mu \\
&\quad+ (\mathcal{D}_{OO}^{(2)})^{\alpha\beta ij\mu\nu}\partial_\mu\partial_\nu+...
\end{split}
\end{align}
where the $...$ denote higher derivatives. The operators $\bar{\mathcal{D}}$ are called non-dissipative and when inserted into \eqref{eq:second_law_thermodynamic_density_frame} yield a solution with $\Delta = 0$ and $N^\mu_{\text{nhs}} = \Theta^\mu_{\text{nhs}}$. In particular, this requires that the non-dissipative non-hydrostatic constitutive relations must vanish, up to a total derivative term, when inserted into the off-shell second law,
\begin{align}
\bar{\mathcal{D}}_{OO}^{\alpha\beta ij}\delta_{\mathcal{B}}B_{\alpha i}\delta_{\mathcal{B}} B_{\beta j} = \partial_\mu \bar{\Theta}^{\mu}_{OO}
\end{align}
and likewise for $\bar{\mathcal{D}}_{\phi\phi}$ and $\bar{\mathcal{D}}_{\psi\psi}$ leading to $\Theta^{\mu}_{\text{nhs,non-diss}} = \bar{\Theta}^{\mu}_{OO}+\bar{\Theta}^{\mu}_{\phi\phi}+\bar{\Theta}^{\mu}_{\psi\psi}$. To leading order in derivatives $\bar{\Theta}^\mu_{OO}=\bar{\Theta}^{\mu}_{\phi\phi} = \bar{\Theta}^{\mu}_{\psi\psi} = 0$ and the matrix $(\bar{\mathcal{D}}_{OO}^{(0)})^{\alpha\beta ij}$ is antisymmetric under $\alpha i \leftrightarrow \beta j$, the matrix $(\bar{\mathcal{D}}_{\phi\phi}^{(0)})^{ab}$ is antisymmetric under $a \leftrightarrow b$ and likewise $(\bar{\mathcal{D}}_{\psi\psi}^{(0)})^{IJ}$ is antisymmetric under $I\leftrightarrow J$. 

The unbarred operators $\mathcal{D}$ produce entropy and are termed dissipative. We may write
\begin{align}
\partial_\mu(N^\mu_{\text{nhs,diss}} - \Theta^\mu_{\text{nhs,diss}}) = \Delta/T + \mathcal{S}^{Ai} \mathcal{D}^{ABij}\mathcal{S}^{Bj}
\end{align}
where 
\begin{align}
\mathcal{S}^{Ai} = \delta^{Ai,\alpha j} \delta_{\mathcal{B}}B_{\alpha j}+\delta^{Ai,a}X_{\text{hs}}^a+\delta^{Ai,I}\delta_{\mathcal{B}}\psi_{I}
\end{align}
and we have defined $\mathcal{D}^{ABij}$ in the obvious way, with e.g. $\mathcal{D}^{\alpha\beta ij} = \mathcal{D}_{OO}^{\alpha \beta ij}$ and $\mathcal{D}^{ab} = \mathcal{D}_{\phi\phi}^{ab}$. The solution to this equation has $N^\mu_{\text{nhs,diss}} = \Theta^\mu_{\text{nhs,diss}}$ and $\Delta/T = - \mathcal{S}^{Ai} \mathcal{D}^{ABij} \mathcal{S}^{Bj}$ showing that the symmetric part of $\mathcal{D}^{ABij}$ under $Ai\leftrightarrow Bj$ must be a negative definite Hermitian operator.

As before, the operator $\mathcal{D}^{ABij}$ has a derivative expansion
\begin{align}
\mathcal{D}^{ABij} = (\mathcal{D}^{(0)})^{ABij} + (\mathcal{D}^{(1)})^{ABij\mu}\partial_\mu + ...
\end{align}
As we will explain in more detail shortly, within linear response, because we expand around an equilibrium state with a particular configuration of the fields, it is only the $\mathcal{S}_{Ai}$ which fluctuate and the matrices $(\mathcal{D}^{(n)})^{ABij\mu_1\mu_2...\mu_n}$ can be taken as constants. Since, for 
\begin{align}
N^\mu_{\text{nhs,diss}}-\Theta^\mu_{\text{nhs,diss}} =  \frac{1}{2}(\mathcal{D}^{(1)})^{ABij\mu}\mathcal{S}_{Ai} \mathcal{S}_{Bj}\end{align}
we have
\begin{align}
\partial_\mu(N^\mu_{\text{nhs,diss}}-\Theta^\mu_{\text{nhs,diss}})= (\mathcal{D}^{(1)})^{ABij\mu}\mathcal{S}_{Ai}\partial_\mu(\mathcal{S}_{Bj}) 
\end{align}
such a term would not be sign-definite and hence cannot appear in the expansion of $\mathcal{D}$. Note too that choosing 
\begin{align}
N^\mu_{\text{nhs,diss}}-\Theta^\mu_{\text{nhs,diss}} =  (\mathcal{D}^{(2)})^{ABij\mu\nu}\mathcal{S}_{Ai}\partial_\nu\mathcal{S}_{Bj}
\end{align}
we have
\begin{align}
\begin{split}
&\partial_\mu (N^\mu_{\text{nhs,diss}}-\Theta^\mu_{\text{nhs,diss}}) \\
&= (\mathcal{D}^{(2)})^{ABij\mu\nu}\mathcal{S}_{Ai}\partial_\mu\partial_\nu (\mathcal{S}_{Bj})\\
&\quad+(\mathcal{D}^{(2)})^{ABij\mu\nu}\partial_\mu(\mathcal{S}_{Ai})\partial_\nu (\mathcal{S}_{Bj} )\,.
\end{split}
\end{align}
The first term is seen to cancel the $\mathcal{O}(\partial^2)$ term in the expansion of $\mathcal{D}^{ABij}$ where as the second line is constrained again to be negative definite. Hence, via judicious integration by parts, we can write
\begin{align}
\begin{split}
\partial_\mu \tilde{\Theta}^\mu + \Delta/T &= -\mathcal{S}_{Ai}(\mathcal{D}^{(0)})^{ABij}\mathcal{S}_{Bj} \\
&\quad-\partial_\mu(\mathcal{S}_{Ai})(\mathcal{D}^{(2)})^{ABij\mu\nu}\partial_\nu(\mathcal{S}_{Bj})+...
\end{split}
\end{align}
where $\mathcal{D}^{(0)}$ and $\mathcal{D}^{(2)}$, etc. are constrained to be positive semidefinite. Here we have included a potential $\tilde{\Theta}^\mu$ term which can be used to cancel non-sign definite terms, e.g. $\partial_i(\tilde{\Theta}^i \equiv Q^{(\alpha\beta)i}\delta s_\alpha \delta s_\beta)  = 2Q^{(\alpha\beta)i}\delta s_\alpha \partial_i\delta s_\beta$. We will use such a term below.

However, as currently formulated, the out-of-equilibrium currents are not easily compared to the discussion in the main text. To do so, we note that we can compare to equilibrium with $\delta_{\mathcal{K}} B_{a\mu} = \delta_{\mathcal{K}}\phi_a = \delta_{\mathcal{K}}\psi_I = 0$, leading to the identities
\begin{align}
\begin{split}
\bar{v}^i F^\alpha_{ti} &=  \bar{T}\partial_t\bar{s}_{\alpha} \,,\\
F^{\alpha}_{ti} &= - \bar{v}^jF^{\alpha}_{ji} - \bar{T}\partial_i\bar{s}_\alpha \, , \\
\bar{\xi}_{a t} &= -\bar{v}^j\xi_{aj} + \bar{T}g^a_{\alpha}\bar{s}_{\alpha} \, , \\
\partial_t\bar{\psi}_I &= - \bar{v}^j\partial_j\bar{\psi}_I \,.
\end{split}
\end{align}
Then, defining the material derivative $\hat{D}(\cdot) \equiv \partial_t(\cdot)+\bar{v}^i\partial_i(\cdot) = 0$ and
\begin{align}
\begin{split}
\delta s^\epsilon &= K^t-\beta^t, \quad
\delta s^{g_i} = \beta^i - K^i, \quad \delta\psi_I=\psi_I-\bar\psi_I\\
\delta s^{\hat{\alpha}} &= \lambda_{\hat{\alpha}}-\Lambda_{\hat{\alpha}} - \delta s^\epsilon B_{\hat{\alpha}t} + \delta s^{g_i} B_{\hat{\alpha}i}
\end{split}
\end{align}
we have
\begin{align}
\begin{split}
\delta_{\mathcal{B}} B_{\alpha i} &= -\delta s^{\epsilon}F^{\alpha}_{t i} + \delta s^{g_j}F^{\alpha}_{ji} + \partial_i \delta s_{\alpha} \,\\
\delta_{\mathcal{B}} \psi_I &= -\delta s^{\epsilon}\partial_t\bar{\psi}_I + \delta s^{g_i}\partial_i \bar{\psi}_I + \frac{1}{\bar{T}}\hat{D}\delta\psi_I.
\end{split}
\end{align}
Note that, because there is no equation of motion for $\delta\psi_I$ we should set $\delta \psi_I = 0$. In other words, fluctuations in symmetry breaking fields can be repackaged in terms of fluctuations of the hydrodynamic fields.

To simplify future results, we can define 
\begin{align}
\begin{split}
L^{Ai}_{g_j} &= \delta^{Ai,\alpha k}F^{\alpha}_{jk} + \delta^{Ai,I}\partial_j\bar{\psi}_I \,,\\
L^{Ai}_{\epsilon} &= -\delta^{Ai,\beta j}F^{\beta}_{tj} - \delta^{Ai,I}\partial_t\bar{\psi}_I \,\\
\tilde{s}_{Ai} &= \delta^{Ai,\beta j}\partial_j\delta s_\beta + \delta^{Ai,a}X^a_{\text{hs}} + \delta^{Ai,I}\hat{D}\delta\psi_I
\end{split}
\end{align}
so that
\begin{align}
\mathcal{S}_{Ai} = L^{Ai}_\alpha \delta s_{\alpha} + \tilde{s}_{Ai}\,.
\end{align}
We also define
\begin{align}
\mathcal{J}_{\text{nhs}}^{Ai} = \delta^{Ai,\alpha j}j^{\alpha j}_{\text{nhs}} + \delta^{Ai,a}\delta_{\mathcal{B}}\phi_a+\delta^{Ai,I}\Psi^I_{\text{nhs}}\,.
\end{align}\

 So, we may rewrite the out-of-equilibrium second law as $-\partial_\mu\tilde{\Theta}^\mu-\Delta/T = J^{Ai}_{\text{nhs}}\delta_{\mathcal{B}}S_{Ai} = \mathcal{J}^{Ai}_{\text{nhs}}\mathcal{S}_{Ai}$,
\begin{align}
\label{eq:out_of_equilibrium_2nd}
\begin{split}
-\partial_\mu\tilde{\Theta}^\mu-\Delta/T=  f_{\text{nhs}}^\alpha \delta s_{\alpha} + j^{\alpha i}_{\text{nhs}}\partial_i \delta s_{\alpha}  + X_{\text{hs}}^{a}\delta_{\mathcal{B}}\phi_a 
\end{split}
\end{align}
where $f^\alpha_{\text{nhs}} = \mathcal{J}^{Ai}_{\text{nhs}}L^{Ai}_{\alpha}$,
\begin{align}
\begin{split}
\label{eq:tilde_to_nhs}
f_{\text{nhs}}^{g_j} &= F^{\alpha}_{ji}j^{\alpha i}_{\text{nhs}} + \Psi^I_{\text{nhs}}\partial_j\bar{\psi}_I , \\
f_{\text{nhs}}^{\epsilon} &= -(F^{\alpha}_{ti}j^{\alpha i}_{\text{nhs}} + \Psi^I_{\text{nhs}}\partial_t\bar{\psi}_I) \, ,
\end{split}
\end{align}
In condensed notation,
\begin{align}
\begin{split}
f^{\alpha}_{\text{nhs}} &= L^{Ai}_{\alpha}(\mathcal{D}+\bar{\mathcal{D}})^{ABij}\mathcal{S}_{Bj} \,\\
&=L^{Ai}_{\alpha}(\mathcal{D}+\bar{\mathcal{D}})^{ABij}L^{Bj}_\beta\delta s_\beta + L^{Ai}_{\alpha}(\mathcal{D}+\bar{\mathcal{D}})^{ABij}\tilde{s}_{Bj}\, ,\\
&\equiv -\tilde{Q}_0^{\alpha\beta}\delta s_\beta - (Q_1+\tilde{N}_0)^{\beta i}_\alpha\partial_i\delta s_\beta- (Y_1+\bar{Y}_1)^{\alpha a}X^a_{\text{hs}},
\end{split}
\end{align}
\begin{align}
\begin{split}
j^{\alpha i}_{\text{nhs}} &= (\mathcal{D}+\bar{\mathcal{D}})^{\alpha Bij}\mathcal{S}_{Bj}\\
&=(\mathcal{D}+\bar{\mathcal{D}})^{\alpha Bij}L^{Bj}_{\beta}\delta s_\beta + (\mathcal{D}+\bar{\mathcal{D}})^{\alpha Bij}\tilde{s}_{Bj} \,\\
&\equiv -(Q_1-\tilde{N}_0)^{\alpha i}_\beta\delta s_\beta - \tilde{N}_1^{\alpha \beta ij} \partial_j\delta s_{\beta} - (Z_1+\bar{Z}_1)^{\alpha a i}X^a_{\text{hs}} \,,
\end{split}
\end{align}
and
\begin{align}
\begin{split}
\delta_{\mathcal{B}}\phi_a &= \sigma_\phi^{ab}X^b_{\text{hs}}-(Y_1-\bar{Y}_1)^{\alpha a}\delta s_\alpha - (Z_1-\bar{Z}_1)^{\alpha a i}\partial_i\delta s_\alpha\,.
\end{split}
\end{align}

where
\begin{align}
\begin{split}
(Q_1)^{\beta i}_{\alpha} &= -\mathcal{D}^{\beta Bij}L_\alpha^{Bj}, \quad (\tilde{N}_0)^{\beta i}_{\alpha} =\bar{\mathcal{D}}^{\beta Bij}L_\alpha^{Bj} \,,\\
(Y_1)^{\alpha a} &= -\mathcal{D}^{aBj}L^{Bj}_{\alpha}, \quad\;\;\; (\bar{Y}_1)^a_{\alpha} = \bar{\mathcal{D}}^{aBj}L_\alpha^{Bj}\,,\\
(Z_1)^{\alpha a i} &= -\mathcal{D}^{\alpha a i} \,, \quad\quad\;\;\;(\bar{Z}_1)^{\alpha a i} = \mathcal{D}^{\alpha a i} \,,\\
&\tilde{Q}_0^{\alpha\beta}  = -L_\alpha^{Ai}(\mathcal{D}+\bar{\mathcal{D}})^{ABij}L_\beta^{Bj}\, ,\\
 &\tilde{N}_1^{\alpha\beta ij} = -(\mathcal{D}+\bar{\mathcal{D}})^{\alpha\beta ij} \,.
\end{split}
\end{align}

By defining $\tilde{\Theta}^t = 0$ and using that $X^a_{\text{hs}} = \partial_i X^{ai}_{\text{hs}}$ we can define
\begin{align}
\begin{split}
-\tilde{\Theta}^i &= 2[Q_1]^{\;\;\alpha) i}_{(\beta} \delta s_{\alpha}\delta s_\beta 
\end{split}
\end{align}
to cancel the $Q_1$ contribution to $f^\alpha_{\text{nhs}}$, and $j^{\alpha i}_{\text{nhs}}$.

Ultimately, then
\begin{align}
\Delta/T = \left(\begin{array}{c}\delta s_\alpha\\ X^a_{\text{hs}} \\\partial_i \delta s_{\alpha} \end{array}\right)^T
\mathfrak{D}
\left(\begin{array}{c}\delta s_\beta \\ X^b_{\text{hs}} \\\partial_j \delta s_\beta \end{array}\right) +...\geq 0
\end{align} 
where the $...$ denotes higher derivative terms and
\begin{align}
\mathfrak{D} = \left(\begin{array}{ccc} \tilde{Q}_0^{\alpha\beta} & Y_1^{\alpha b} & 0\\
Y_1^{\beta a} & -\sigma^{ab}_{\phi} & Z_1^{\beta a j}  \\
0 & Z_1^{\alpha b i}& \tilde{N}_1^{\alpha \beta ij} \end{array}\right) \leq 0 \,
\end{align}
which requires that $\tilde{Q}_0, -\sigma_\phi, \tilde{N}_1$ are all positive definite square matrices and that 
\begin{align}
\left(\begin{array}{cc} \tilde{Q}_0^{\alpha\beta} & Y_1^{\alpha b} \\ Y_1^{\beta a} & -\sigma_\phi^{ab}\end{array}\right)\leq 0, 
\quad \left(\begin{array}{cc} -\sigma_\phi^{ab} & Z_1^{\beta a j}\\Z_1^{\alpha b i} & \tilde{N}_1^{\alpha\beta ij} \end{array}\right) \leq 0\,.
\end{align}
Unfortunately, the $Y_1$ terms pose an issue for our stability argument because they cannot be constrained on their own. However, we may express $X^a_{\text{hs}} = \partial_i (\frac{h^{ai}}{T})$ which follows from the fact that the free energy functional must be a function of $\partial_i\phi_a$ and always includes a term proportional to $(\partial\phi_a)^2$. Then, including a term
\begin{align}
\tilde{\Theta}^i\to \tilde{\Theta}^i - 2Y_1^{\alpha a}\delta s_\alpha \frac{h^{ai}}{T}
\end{align}
eliminates the $Y_1$ terms from $\delta_{\mathcal{B}}\phi_a$ and $f^\alpha_{\text{nhs}}$ but it introduces a term
\begin{align}
\partial_ij^{\alpha i}_{\text{nhs}} \to \partial_i j^{\alpha i}_{\text{nhs}} + Y_1^{\alpha a}X^{a}_{\text{hs}} \sim i\mathfrak{v}q \,.
\end{align}
where we used that $X^a_{\text{hs}}\propto iq$ this leads to an extra real propagating component (that is not hydrostatic!) in the dispersion relations and does not need to be sign constrained for stability. This is a mode which comes from a combination of explicit and spontaneous symmetry breaking.

\section{\label{app:hydrostatic_corrections} Hydrostatic corrections}
In this appendix, we give an example of a hydrostatic correction to show how its role in hydrodynamic fluctuations is to modify the eigenvector of \eqref{linearized_eom} but not the dispersion relation.

For simplicity, we will consider a neutral fluid with conserved energy density, $\epsilon$, and conserved momentum, $g^i$. The conjugate sources are $-1/T$ and $v^i/T$ respectively, i.e. in the ideal limit, we have the familiar relations
\begin{align}
\begin{split}
ds_{\text{id}} &= (1/T)d\epsilon_{\text{id}} - (v^i/T)dg_{\text{id}}^i \,\\
d(p_{\text{id}}/T) &= -\epsilon_{\text{id}} d(1/T) + g_{\text{id}}^i d(v^i/T)\,\\
\epsilon_{\text{id}} + p_{\text{id}} &= Ts_{\text{id}}  + v^i g_{\text{id}}^i.
\end{split} 
\end{align}
These relations can also be found via an ideal order generating functional $\mathcal{F}_{\text{id}} = -p_{\text{id}}$. Now, consider the following one derivative correction in the hydrostatic generating functional, $\mathcal{F}/T = \mathcal{F}_{\text{id}}/T + \mathcal{F}_1/T$, where
\begin{align}
\mathcal{F}_1/T =  -\gamma (v^i/T)\partial_i (-1/T)
\end{align}
where $\gamma$ is a function of $v^2\equiv v_i v^i$ and $T$. Such a term leads to a correction to the momentum and energy densities, 
\begin{align}
\begin{split}
d(\mathcal{F}_1/T) = -\epsilon_1d\left(-\frac{1}{T}\right)-g^i_1d\left(\frac{v^i}{T}\right) + \partial_i d\theta_1^i
\end{split}
\end{align}
where
$\epsilon = \epsilon_{\text{id}}+\epsilon_1$ and $g^i=g^i_{\text{id}}+g^i_1$ and
\begin{align}
\begin{split}
\label{densities_hydrostaticcorrections}
\epsilon_1&= -E^{ij}\partial_i\left(\frac{v_j}{T}\right) \, \quad
g_1^i =E^{ij}\partial_j\left(-\frac{1}{T}\right)  \,, \\
&\quad E^{ij} = \left(\gamma\delta^{ij} +2\frac{\partial\gamma}{\partial v^2} v^i v^j\right) \,,\\
&\quad d\theta_1^i = -\gamma \frac{v^i}{T}d\left(-\frac{1}{T}\right) \,.
\end{split}
\end{align}
The hydrostatic correction to the generating functional leads to a pressure $p = p_{\text{id}}+p_1$ with $p_1 = -\mathcal{F}_1 + T\partial_i\theta^i$ and we can define a correction to the entropy via
\begin{align}
s_1 \equiv (\epsilon_1 + p_1 - v^ig_1^i)/T
\end{align}
so that we can continue to write the entropy flux as
\begin{align}
j_s^i = -s_A j^{Ai} + p \frac{v^i}{T}
\end{align}
and the energy and momentum fluxes as
\begin{align}
\begin{split}
j_\epsilon^i &= (p_{\text{id}}+\epsilon_{\text{id}})v^i + (p_1+\epsilon_1)v^i \, ,\\
\tau^{ij} &= (p_{\text{id}}+p_1)\delta^{ij} + (g_{\text{id}}^i + g_1^{i})v^j \,.
\end{split}
\end{align}
Note that although $p_1$ is implicitly defined in terms of $\theta^i$, we don't need an expression for this function within linear response, only its variations which are given in \eqref{densities_hydrostaticcorrections}.

Next, considering linearized fluctuations, $s^\epsilon = -\frac{1}{\bar{T}} + \delta s^\epsilon e^{-i\omega t+ iq_j x^j}$, $s^{g^i} = \frac{\bar{v}^i}{\bar{T}} + \delta s^{g^i} e^{-i\omega t + iq_j x^j}$, $\epsilon = \bar{\epsilon}+(\delta\epsilon_{\text{id}}+\delta\epsilon_1)e^{-i\omega t + iq_j x^j}$, and $g^i = \bar{g}^i + (\delta g^i_{\text{id}}+\delta g_1^i)e^{-i\omega t + iq_j x^j}$, so that
\begin{align}
\begin{split}
\delta g_1^i &= \left(\frac{\partial E^{ij}}{\partial s^{g^k}}\partial_j\left[-\frac{1}{\bar{T}}\right]\right)\delta s^{g^k} \\
&\quad+\left(\frac{\partial E^{ij}}{\partial s^{\epsilon}}\partial_j\left[-\frac{1}{\bar{T}}\right]+ iq_j \bar{E}^{ij}\right)\delta s^\epsilon \,\\
\delta \epsilon_1 &= -\left(\frac{\partial E^{ij}}{\partial s^\epsilon}\partial_i\left[\frac{\bar{v}^j}{\bar{T}}\right]\right)\delta s^\epsilon\\
&\quad -\left(\frac{\partial E^{ik}}{\partial s^{g^j}}\partial_i\left[\frac{\bar{v}^k}{\bar{T}}\right]+ iq_i\bar{E}^{ij}\right) \delta s^{g^j} \,.
\end{split}
\end{align}
For completeness,
\begin{align}
\begin{split}
\frac{\partial E^{ij}}{\partial s^\epsilon} &=\bar{T} \left(\bar{T}\frac{\partial \gamma}{\partial T}+2\bar{v}^2\frac{\partial \gamma}{\partial v^2}\right) \delta^{ij}\\
&+ 2\bar{v}^i\bar{v}^j \bar{T}\left(2\frac{\partial \gamma}{\partial v^2}+\bar{T}\frac{\partial^2\gamma}{\partial v^2\partial T}+2\bar{v}^2\frac{\partial^2\gamma}{(\partial v^2)^2}\right) \,,\\
\frac{\partial E^{ij}}{\partial s^{g^k}} &= 2\bar{T}\frac{\partial \gamma}{\partial v^2} \left(\bar{v}^k\delta^{ij}+\bar{v}^i\delta^{jk}+\bar{v}^j\delta^{ik}\right)\\
&\quad+4\bar{T}\bar{v}^i\bar{v}^j\bar{v}^k\frac{\partial^2 \gamma}{(\partial v^2)^2}\,.
\end{split}
\end{align} 
It is observed that the $\mathcal{O}^q$ terms are fully symmetric in their spacetime indices. If we were to impose Onsager symmetries, we could partially fix the functional dependence of $\gamma$ on $T$ and $v^2$, but we will avoid doing so for now. To simplify future expressions, we write 
\begin{align}
\delta O^A = \chi^{AB}\delta s^B = (\chi_0^{AB}+iq_k \chi_1^{ABk}+...)\delta s^B
\end{align}
where
\begin{align}
\begin{split}
\delta \epsilon &= \left(\chi_0^{\epsilon\epsilon}+iq_k\chi_1^{\epsilon\epsilon k}\right)ds^\epsilon +  \left(\chi_0^{\epsilon g^j}+iq_k\chi_1^{\epsilon g^j k}\right)ds^{g^j} \,, \\
\delta g^i &=  \left(\chi_0^{g^i\epsilon}+iq_k\chi_1^{g^i\epsilon k}\right)ds^\epsilon+ \left(\chi_0^{g^i g^j}+iq_k\chi_1^{g^i g^j k}\right)ds^{g^j} \,.
\end{split}
\end{align}
where $\chi_0^{AB}$ includes both ideal and hydrostatic corrections and $\chi_1^{AB}$ contains only the hydrostatic correction. For instance,
\begin{align}
\chi_0^{\epsilon\epsilon} = \frac{\partial \epsilon_{\text{id}}}{\partial (-1/T)} - \frac{\partial E^{ij}}{\partial (-1/T)}\partial_i \left(\frac{\bar{v}^j}{\bar{T}}\right).
\end{align}
Note that in the hydrostatic limit of $|q|\ll 1$, we can invert $\chi^{AB}$ to linear order
\begin{align}
(\chi^{-1})^{AB} = (\chi_0^{-1})^{AB} - iq_k (\chi_0^{-1}\cdot \chi_1\cdot \chi_0^{-1})^{ABk}+\mathcal{O}(q^2)
\end{align}

In order to fully specify the hydrodynamic fluctuations, we also need to specify the out-of-equilibrium fluxes. While this is straightforward formally (see the next Section), actual expressions depend on whether or not $\partial_i \bar{v}^j$, $\partial_i \bar{T}$, and $\bar{v}^j$ vanish or not. For the sake of not overcomplicating the analysis, we will set $\partial_i\bar{T} = \partial_i \bar{v}_j = \bar{v}^j = 0$ for the rest of the analysis. Note that in this case, hydrostatic corrections to $\chi_0^{AB}$ vanish, but
\begin{align}
\chi_1^{ABi} = \gamma \delta^{\epsilon [A }\delta^{B] g^i} .
\end{align} 

For the case of a neutral fluid, working in a hydrodynamic frame in which the densities do not receive corrections, and one in which $\bar{v}^i \neq 0$,
\begin{align}
\begin{split}
\tilde{j}^i_\epsilon &= \kappa \partial_i \left(-\frac{1}{T}\right) \, ,\\
\tilde{\tau}^{ij} &= \zeta \partial_k (\frac{v^k}{T})\delta^{ij} + \frac{\eta}{2}\left(\partial_i [\frac{v^j}{T}] + \partial_j [\frac{v^i}{T}] - \frac{2}{d}(\partial\cdot [\frac{v}{T}])\delta^{ij}\right)\,.
\end{split}
\end{align}
with constraints from the second law that $\kappa \geq 0, \zeta \geq 0, \eta \geq 0$.

At this point, we are ready to find the dispersion relations of the hydrodynamic modes. The linearized equations of motion in the longitudinal sector read
\begin{align}
\begin{split}
0&=-i\omega \left(\begin{array}{cc}
\chi_0^{\epsilon\epsilon} & iq\gamma \\
-iq\gamma & \chi_0^{gg}\end{array}\right)\left(\begin{array}{c}\delta s^\epsilon(q) \\ \delta s^g(q)\end{array}\right) 
\\
&+ i q \left(\begin{array}{cc} iq\kappa & (\bar{p}+\bar{\epsilon})\bar{T} \\ (\bar{p}+\bar{\epsilon}) \bar{T} &   iq (\zeta+ \frac{d-1}{d}\eta) \end{array}\right)\left(\begin{array}{c}\delta s^\epsilon(q) \\ \delta s^g(q)\end{array}\right)
\end{split}
\end{align}
The solutions to this equation are
\begin{align}
\begin{split}
\omega_\pm(q) &= \pm\frac{\bar{T}(\bar{\epsilon}+\bar{p})}{\sqrt{\chi_0^{\epsilon\epsilon}\chi_0^{gg}}}q - i\frac{\kappa\chi_0^{gg} +\chi_0^{\epsilon\epsilon} (\zeta + \frac{d-1}{d}\eta)}{2\chi_0^{\epsilon\epsilon}\chi_0^{gg}}q^2 + ...
\end{split}
\end{align}
corresponding to eigenvectors
\begin{align}
\begin{split}
&\left(\begin{array}{c} 
\delta s^\epsilon(q) \\
\delta s^g(q)
\end{array}\right)_\pm \\
&= 
\left(\begin{array}{c}
\mp\sqrt{\frac{\chi_{0}^{gg}}{\chi_0^{\epsilon\epsilon}}} - iq \frac{(\zeta+\frac{d-1}{d}\eta)\chi_0^{\epsilon\epsilon}-\kappa \chi_0^{gg}}{2\bar{T}\chi_0^{\epsilon\epsilon}(\bar{\epsilon}+\bar{p})} - iq\frac{\gamma}{\chi_0^{\epsilon\epsilon}}+... \\ 1 
\end{array}\right)
\end{split}
\end{align}
The hydrostatic correction $\gamma$ does not affect the spectrum $\omega(q)$, but it does correct the eigenvectors themselves at $\mathcal{O}(q)$. 

There are no changes to the transverse sector from the hydrostatic correction.

\section{Hydrostatics when $\partial_\mu \bar{s}_A \neq 0$}
We always require that $\partial_\mu(1/\bar{T})$ and $\partial_\mu(\bar{v}^\nu)$ vanish, since relaxing this requirement means that we are working in curved spacetime. In particular, our results for the hydrostatic constitutive relations tell us that (setting $\ell_{eb} =1$ for now)
\begin{align}
\hat{D}\bar{O}^A +\bar{\frac{\delta v^\mu}{\delta s_A}}\partial_\mu(\frac{\bar{p}}{\bar{T}})=-\frac{\bar{O}^C}{\bar{T}}\bar{v}^{\mu}F^C_{[\mu\nu]}\bar{\frac{\delta v^\nu}{\delta s_A}}=\bar{O}^C\partial_\nu\bar{s}_{C} \bar{\frac{\delta v^\nu}{\delta s_A}}
\end{align}
consistent with
\begin{align}
\hat{D}\bar{O}^A = 0 \quad\text{and}\quad dp = O^A d_{A}
\end{align}
so that $\bar{O}^A$ is a scalar under diffeomorphisms and gauge transformations. When we linearize the equations of motion, we use that
\begin{align}
\delta O^A = \chi^{AB}\delta s_B
\end{align}
where it may be the case that $\partial_\mu \chi^{AB} \neq 0$. Of course, since $\chi^{AB}$ is effectively $d^2p/ds^Ads^B$, this is only possible in the presence of a background field. Expanding about equilibrium, we have $s_{A} = \bar{s}_A+\delta s_{A}e^{-i\omega t+iq_j x^j}$. Since we care about the impact of inhomogeneous terms, we can take the limit $q\to 0$. Then, using
\begin{align}
\partial_i(\delta p) = \partial_i(\bar{O}^A\delta s_A) = \delta s^A\partial_i\bar{O}^A
\end{align}
we have
\begin{align}
\begin{split}
0&=-i\omega \delta O^A +\left(\chi^{AB}\partial_\mu\bar{s}_B\right)\delta v^\mu+\left[\left(\bar{\frac{\delta v^\mu}{\delta s^A}}\partial_\mu\bar{O}^B\right)\delta s_B\right]\\
&\quad+\cancel{ \left(\delta^{A\epsilon}\partial_i \bar{p}\right)\delta v^i}- \cancel{\left(\delta^{A\epsilon} \partial_i \bar{p}\right)\delta v^i}\\
&\quad  + \left(\frac{\bar{O}^C}{\bar{T}}F^C_{[\mu\nu]}\bar{\frac{\delta v^\nu}{\delta s^A}}\bar{\frac{\delta v^\mu}{\delta s^B}}\right)\delta s_B\\
&\quad-\left[\left(\bar{\frac{\delta v^\nu}{\delta s^A}}\partial_\nu\bar{s}_C\right)\delta O^C\right]
\end{split}
\end{align}
Now, since $\partial_\mu \bar{O}^B = \chi^{AB}\partial_\mu\bar{s}_B$ and $\delta O^C = \chi^{CB}\delta s_B$, the terms in square brackets combine to give
\begin{align}
-2\frac{\delta v^\mu}{\delta s^A}\partial_\mu \bar{s}_B\chi^{[BC]}\delta s_C\,.
\end{align}
Since $\chi^{AB}\sim \delta^2p/\delta s^A\delta s^B$ we expect it to be symmetric and this vanishes.

When $A \neq g_i,\epsilon$, there exists a solution for any $\omega$ such that for $B \neq g_i, \epsilon$
\begin{align}
\delta s^B = -\frac{i}{\omega}\delta v^\mu\partial_\mu\bar{s}_B - (\chi^{-1})^{BC}\left[\chi^{C\epsilon}\delta s^\epsilon+\chi^{Cg_i}\delta s^{g_i}\right]
\end{align}
This can be used to eliminate the $\delta s_B$ from the equations of motion for $\delta \epsilon$ and $\delta g_i$. Denoting indices which don't span $\epsilon, g_i$ with a hat and those that do with a dot
\begin{align}
\begin{split}
0&=-i\omega \dot{\chi}^{\dot{A}\dot{B}}\delta s_{\dot{B}} + \left(\frac{\bar{O}^C}{\bar{T}}F^C_{[\mu\nu]}\frac{\delta v^\nu}{\delta s^{\dot{A}}}\frac{\delta v^\mu}{\delta s^{\dot{B}}}\right)\delta s_{\dot{B}} 
\end{split}
\end{align}
where we defined
\begin{align}
\dot{\chi}^{\dot{A}\dot{B}} = \left(\chi^{\dot{A}\dot{B}}-\chi^{\dot{A}\hat{C}}(\chi^{-1})^{\hat{C}\hat{D}}\chi^{\hat{D}\dot{B}}\right)\,.
\end{align}
Using block inversion on the matrix $\chi^{AB}$, we see that $(\chi^{-1})^{\dot{A}\dot{B}} = (\dot{\chi}^{-1})^{\dot{A}\dot{B}}$ and so it follows that since $\chi^{AB}$ is positive definite, so too is $\dot{\chi}^{\dot{A}\dot{B}}$. This demonstrates that at $\ell_{eb}$, the inhomogeneous term appearing in the linearized equations of motion is 
\begin{align}
Q^{[\dot{A}\dot{B}]} = \frac{\bar{O}^C}{\bar{T}}F^C_{[\mu\nu]}\frac{\delta v^\nu}{\delta s^{\dot{A}}}\frac{\delta v^\mu}{\delta s^{\dot{B}}}
\end{align}
is antisymmetric, as emphasized in the main text.


\section{More specific constraints on the hydrodynamic dispersion relations}
Note that our argument for stability is generic, applying to the hydrodynamic modes for all values of $q$ and $\ell_{eb}$ within the hydrodynamic regime. At the same time, this does not tell us about the polynomial expansion of the spectrum in terms of $q$ and $\ell_{eb}$. In fact, when $\chi_0^{-1}$ is positive definite, each of these coefficients can be constrained as well by writing
\begin{align}
\begin{split}
M^{AB} &=\sum_{j,k=0}\ell_{eb}^j(iq)^kM_{j,k}^{CB} \,\\
\tilde{M}^{AB}(q,\ell_{eb}) &= \sum_{j,k=0}\ell_{eb}^j(iq)^k\left([\chi_0^{-1}\cdot M_{j,k}]^{CB}+\tilde{M}^{CB}_{j,k}\right).
\end{split}
\end{align}
Then the eigenvalues have an expansion
\begin{align}
\tilde{m}^{(A)}(q,\ell_{eb})=\sum_{j,k=0}\ell_{eb}^j(iq)^k\tilde{m}_{j,k}^{(A)}
\end{align}
where $\tilde{m}_{j,k}$ are the eigenvalues of $(\chi_0^{-1}\cdot M_{j,k})^{AB}$. The remaining terms in the expansion for $\tilde{M}^{AB}\delta s^{B}(q,\ell_{eb})$, characterized by the matrices $\tilde{M}_{j,k}$ arise from the terms in expansion of $\chi^{-1}$ with $\chi_{>0}^{-1}$, and hence contain only the matrices $M_{i<j, l<k}$. They are cancelled in the linearized equations of motion by terms in the expansion of $\delta s^{B}(q,\ell_{eb})$ that appear at the same order. For instance, we may write the expansion of the eigenvector corresponding to eigenvalue $\tilde{m}^{(A)}(q)$ as
\begin{align}
\delta s_{(A)}^{B}(q,\ell_{eb}) = \sum_{j,k=0}\ell_{eb}^j(iq)^k\delta s_{(A),j,k}^B\,.
\end{align}
Then, if the lowest order non-vanishing contribution to the eigenvalue $\tilde{m}^{(A)}$ is $\tilde{m}^{(A)}_{a,b}$, we have
\begin{align}
\begin{split}
0&= \sum_{k=0}^i\sum_{l=0}^j \biggl(\chi_0^{-1}\cdot M_{a+i-k,b+j-l}-m_{a+i-k,b+j-l}^{(A)}\mathbb{1}\\
&\hspace{50pt}+\tilde{M}_{a+i-k,b+j-l}\biggr)^{BC}\delta s^C_{(A),k,l}
\end{split}
\end{align}
For any choice of $i,j$ this leads to a solution for $\delta s^B_{(A),i,j}$ in terms of a linear combination $\delta s^{B}_{(A),k<i, l<j}$ and there is a unique choice for $\tilde{m}^{(A)}_{j,k}$.

 In the end, since the $M_{j,k}^{AB}$ have similar symmetry and positivity properties as the matrix $M^{AB}$, the $\tilde{m}^{(A)}_{j,k}$ can be constrained in the same manner. For instance, since the leading term in $\tilde{Q}^{AB}$ is a positive definite matrix of real numbers, the term at $\mathcal{O}(\ell_{eb}^2,q^0)$ in $\omega^{(A)}$ is given by the eigenvalue of $(\chi_0^{-1}\cdot\tilde{Q})^{AB}$, $\tilde{m}_{2,0}^{(A)}$, and its imaginary component is constrained to be non-positive. Since we can do this for all $\tilde{m}^{(A)}_{j,k}$, we can bound the dominant contribution to $\Im \omega^{(A)}$ (i.e. leading in $q$ and $\ell_{eb}$) to be non-positive and the system is linearly stable within the hydrodynamic regime.

\section{Spontaneous condensation in holographic superfluids}
Holographic superfluids are models of large-$N$ strongly-coupled relativistic quantum matter which spontaneously acquire a non-zero expectation value for a global U(1) symmetry breaking operator $\mathcal{O}$ \cite{Gubser:2008px, Hartnoll:2008kx, Hartnoll:2008vx, Sonner:2010yx}. These models are studied via the gauge/gravity duality \cite{Maldacena:1997re, Witten:1998qj} in which dynamics of the quantum system are re-expressed in terms of a weakly-coupled gravitational system in one extra dimension. In the weakly-coupled limit, stochastic and quantum fluctuations are suppressed by the gravitational coupling $G_N \sim 1/N^2 \ll 1$. The hydrodynamic limit of these systems has been studied in depth in \cite{Amado:2009ts, Arean:2021tks, Gouteraux:2022qix, Arean:2023nnn}. 

In addition to charge and energy conservation, the hydrodynamic description of these systems includes the dynamics of the Goldstone, or phase of the order parameter, which arises from the spontaneous symmetry breaking. In contrast, away from the critical point, the norm of the order parameter is typically gapped and does not play a role in the low energy dynamics. However, near the phase transition, the amplitude of the order parameter is parametrically small and becomes a new light degree of freedom. Due to the large-$N$ suppression of fluctuations, this field may be included in the hydrodynamic description \cite{Herzog:2010vz,Donos:2022xfd,Donos:2022qao,Donos:2023ibv}. In this Section, we pursue a preliminary exploration of the impact of this field on superfluid hydrodynamics in the context of thermodynamic instabilities.

We work in bulk spacetime dimension $D=4$ and take $L_{AdS} = 16\pi G_N = 1$ with the standard action for a holographic superfluid \cite{Hartnoll:2008kx, Hartnoll:2008vx, Sonner:2010yx} defined on a manifold $(\Sigma, g)$ with boundary $(\partial\Sigma, \gamma)$,
\begin{align}
\begin{split}
S &= \int_\Sigma d^4x\sqrt{-g}\left(R+6-\frac{F^2}{4}-|D\Psi|^2 +2 |\Psi|^2\right) \\
&\quad+\int_{\partial\Sigma} d^3x\sqrt{-\gamma}\left(2\mathcal{K}-4-|\Psi|^2-R^{(\gamma)}\right) \,.
\end{split}
\end{align}
The first line is the bulk action containing in addition to the Einstein-Hilbert term a U(1) gauge field with field strength $F_{MN} = \partial_M A_N - \partial_N A_M$, a charged scalar $\Psi$ with kinetic term $D_M\Psi = (\partial_M - iqA_M)\Psi$ and mass $m^2 = -2$. Here, $M = \{t,x,y,r\}$ is an index spanning the spacetime dimensions. The second line of the action is a boundary counterterm defined in terms of extrinsic curvature $\mathcal{K}$ and induced boundary metric $\gamma_{\mu\nu}$ where $\mu = \{t,x,y\}$ spans only the boundary spacetime dimensions. The counterterm action is necessary to give a well-defined variational principle \cite{Balasubramanian:1999re, Skenderis:2002wp}.

\begin{figure}[h]
\begin{center}
\includegraphics[scale=.3]{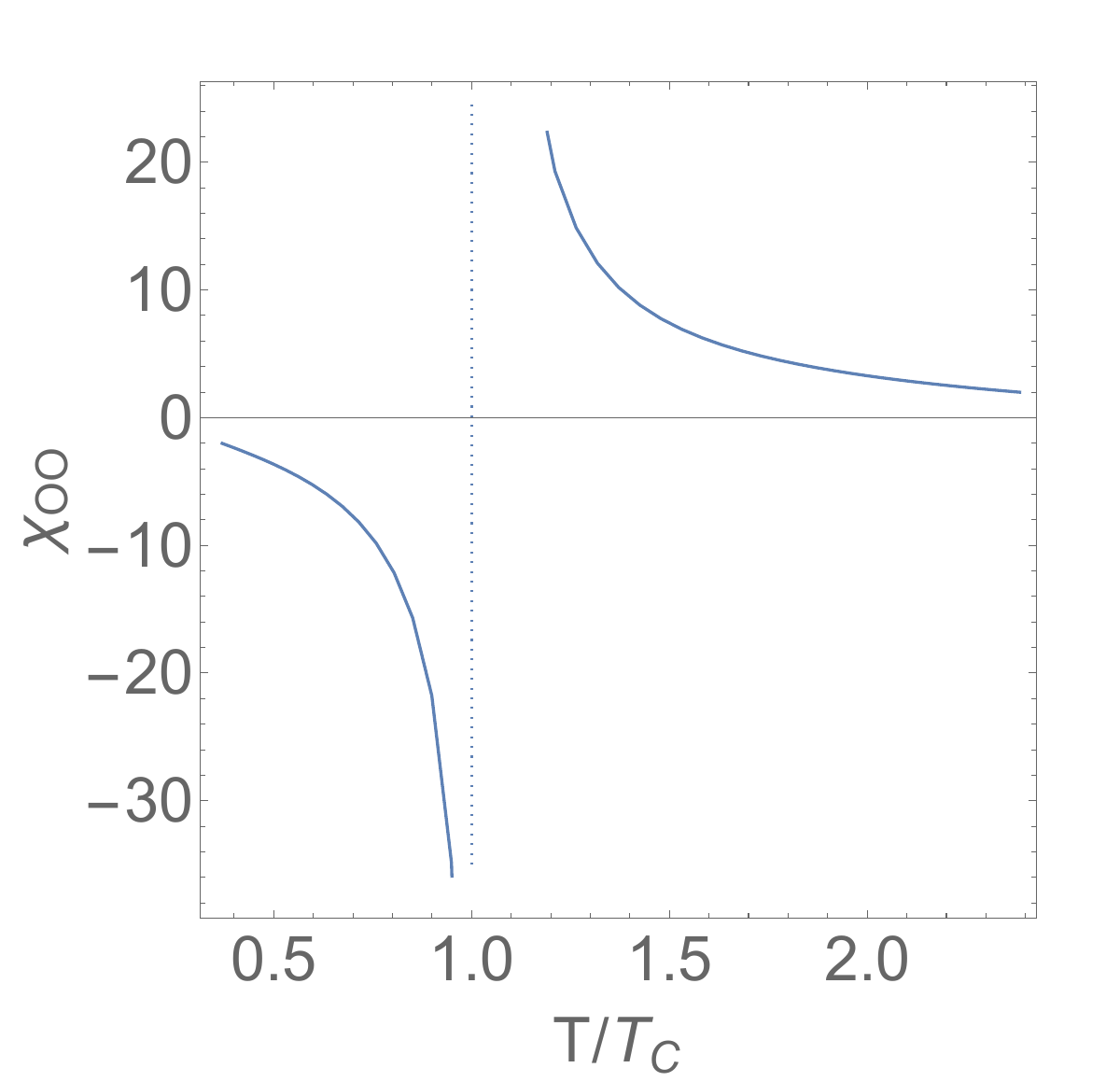}
\caption{\label{decoupledsuperfluidplots} Numerically obtained $\chi_{OO}$ in the decoupled limit case of eq. (\ref{eq:decoupledlimit}) at $s_{\mathcal{O}} = 0$. The critical point for the phase transition is clearly indicated by a diverging $\chi_{\mathcal{O}\mathcal{O}}$ and below $T_c$, $\chi_{OO} < 0$ indicating an instability towards condensation.}
\end{center}
\end{figure}

For the sake of this work, we are interested in the simplest and most direct consequence of a small amplitude for the order parameter: the relation between spontaneous symmetry breaking, the susceptibility $\chi_{\mathcal{O}\mathcal{O}}$ changing sign, and the appearance of a hydrodynamic mode with $\Im \omega > 0$, all at $T_C$. For this reason, we take a simplifying limit in which the scalar field dual to $\mathcal{O}$ decouples from the charge and gravitational equations of motion, i.e. we take
\begin{align}
\label{eq:decoupledlimit}
 |\Psi|\ll 1 \,.
 \end{align}
  In this case, the gravitational and Maxwell fields are given by the Reissner-Nordstr\"om-AdS solution with horizon at $r=r_h$,
\begin{align}
\label{eq:RNlimit}
\begin{split}
ds^2 &= -r^2f(r)dt^2 + \frac{dr^2}{r^2f(r)} + r^2(dx^2+dy^2) \,,\\
f(r) &= 1-\frac{r_h^3}{r^3}-\frac{\mu^2r_h^3}{4r^3}(1-\frac{r_h}{r})\,,
\end{split}
\end{align}
and
\begin{align}
A_t(r) &= \mu(1-\frac{r}{r_h}), \quad A_x=A_y=A_r = 0 \,.
\end{align}
For simplicity, we also set $r_h = q = 1$. The associated temperature of this solution is then $T= \frac{12-\mu^2}{16\pi}$. Finally, we are free to choose the phase of $\Psi$ such that $\Psi = \Psi^* = \psi$. In this simplifying limit, there is one equation to solve
\begin{align}
\label{eq:decoupledEOM}
0 = \frac{1}{\sqrt{-g}}\partial_M(\sqrt{-g}D^M\psi) + iqA^MD_M\psi + 2\psi \,.
\end{align}
In the UV ($r\to \infty$), the charged scalar field has an asymptotic form
\begin{align}
\psi(r) \to \frac{s_{\mathcal{O}}}{r} + \frac{\mathcal{O}}{2r^2} + O(r^{-3}) \,.
\end{align}
Here, $s_{\mathcal{O}}$ is interpreted as the source for the order parameter $\mathcal{O}$. Solutions with $\mathcal{O} \neq 0$ and  $s_{\mathcal{O}} = 0$ indicate spontaneous symmetry breaking. These only occur for sufficiently low $T$ (or large $\mu$) when solutions with $\mathcal{O}\neq 0$ are thermodynamically preferred, or similarly when there is a thermodynamic instability towards condensation. In figure \ref{decoupledsuperfluidplots}, we illustrate this by showing that for the range of $T$ for which condensation is preferred, 
\begin{align}
\label{eq:chiOOdef}
\chi_{\mathcal{O}\mathcal{O}} \equiv  \frac{\partial \mathcal{O}}{\partial s_{\mathcal{O}}}\biggr|_{s_{\mathcal{O}}=0} < 0.
\end{align}
To find this quantity, we solve eq. (\ref{eq:decoupledEOM}) for a range of $s_{\mathcal{O}}$ and numerically compute the derivative, eq. (\ref{eq:chiOOdef}), in the limit $s_{\mathcal{O}}\to 0$.

\begin{figure}[h]
\begin{center}
\includegraphics[scale=.2]{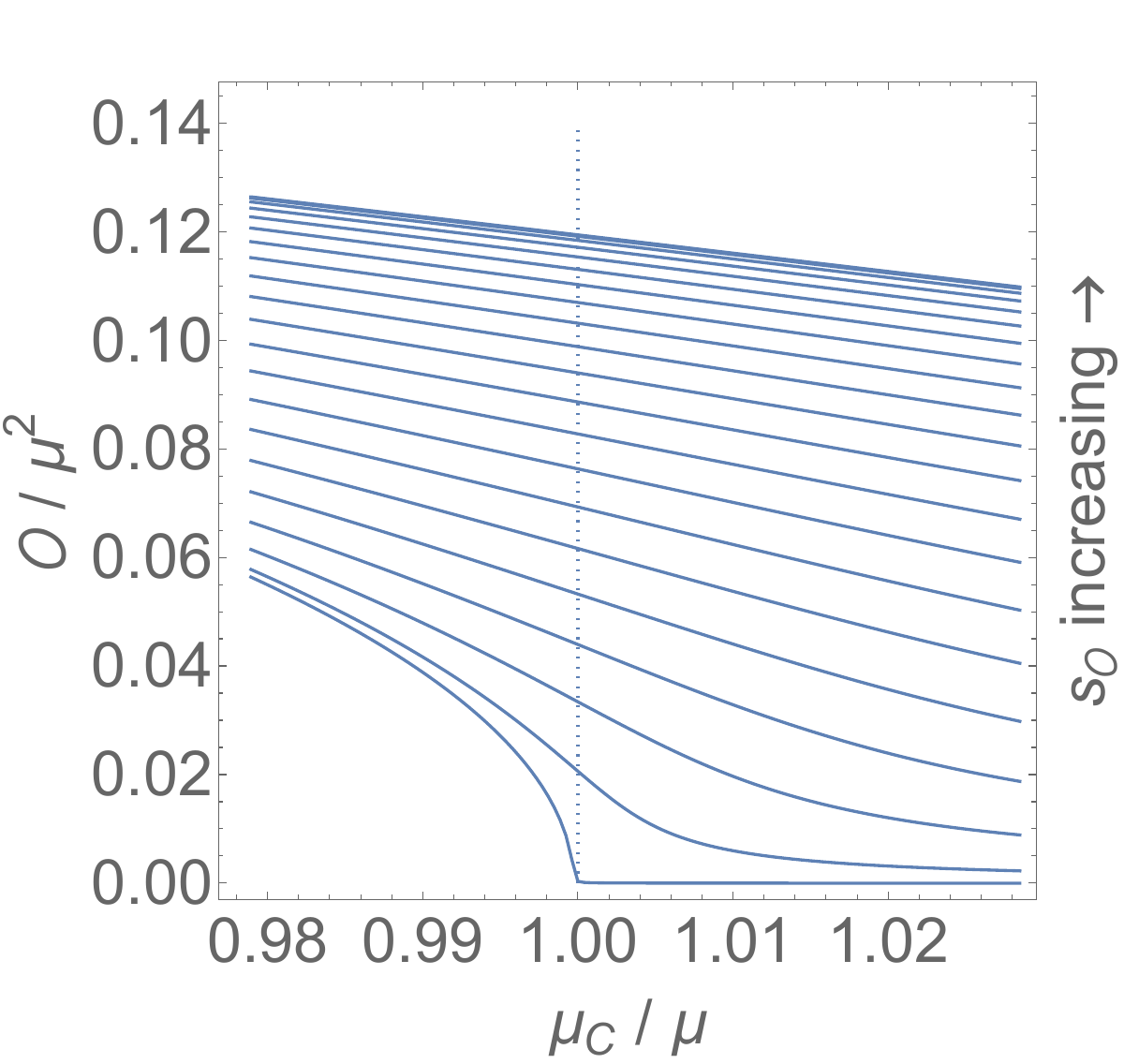}
\includegraphics[scale=.2]{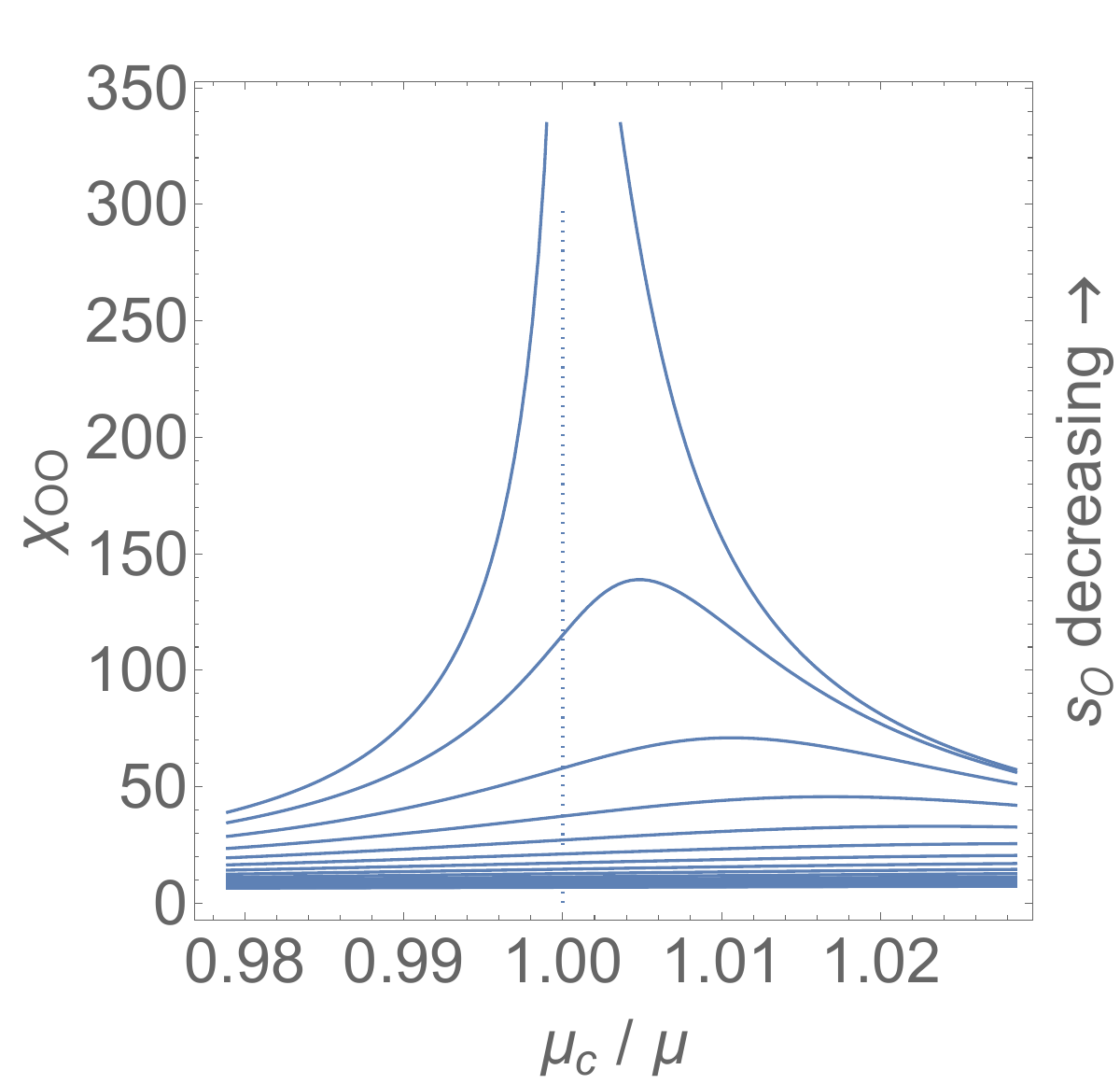}
\caption{\label{probesuperfluidplots} Numerically obtained condensate and $\chi_{OO}$ in the probe limit case of eq. (\ref{eq:probelimit}). On the left, the condensate, on the right, $\chi_{OO}$, with different curves corresponding to $s_{\mathcal{O}}\in [0,1/10]$. In contrast to figure \ref{decoupledsuperfluidplots}, the modified profile for $A_t$ leads to a stable phase with $\mathcal{O}\neq 0$ and $s_{\mathcal{O}} = 0$. Nevertheless, the critical point for the phase transition is clearly indicated by a diverging $\chi_{\mathcal{O}\mathcal{O}}$.}
\end{center}
\end{figure}

The endpoint of this instability is the condensation of the order parameter, which leads to some of the charge density to occupy the superfluid state. A consistent limit solution, then, requires that we take into account charge fluctuations leading to the `probe limit' where we take
\begin{align}
\label{eq:probelimit}
A_M \to A_M/q, \quad \Psi \to \Psi/q
\end{align}
and $q\to \infty$. In other words, we consider coupling charge fluctuations with fluctuations of the order parameter, but decouple all energy dynamics. The analysis is only slightly changed in this limit---we can take $\mu \to 0$ in \eqref{eq:RNlimit} leading to a Schwarzschild-AdS background (with $T=3/4\pi$). Then an additional equation of motion appears for $A_t$
\begin{align}
\label{eq:probeEOM}
0 &= \frac{1}{\sqrt{-g}}\partial_M(\sqrt{-g}F^{Mt}) - 2\psi^2A^t
\end{align}
with UV asymptotics
\begin{align}
A_t \to \mu - \frac{\rho}{r} + O(r^{-2})
\end{align}
and eq. (\ref{eq:decoupledEOM}) is left unchanged. As in the fully decoupled scalar case, for large enough $\mu$, the thermodynamically preferred state has $\mathcal{O}\neq 0$. In this case, however, $A_t$ obtains a modified profile which leads to $\chi_{OO}>0$ in the symmetry broken regime---hence, when some of the charge gets transferred to the superfluid state, the system is thermodynamically stable. We illustrate this in figure \ref{probesuperfluidplots}.

\begin{figure}[h]
\begin{center}
\includegraphics[scale=.3]{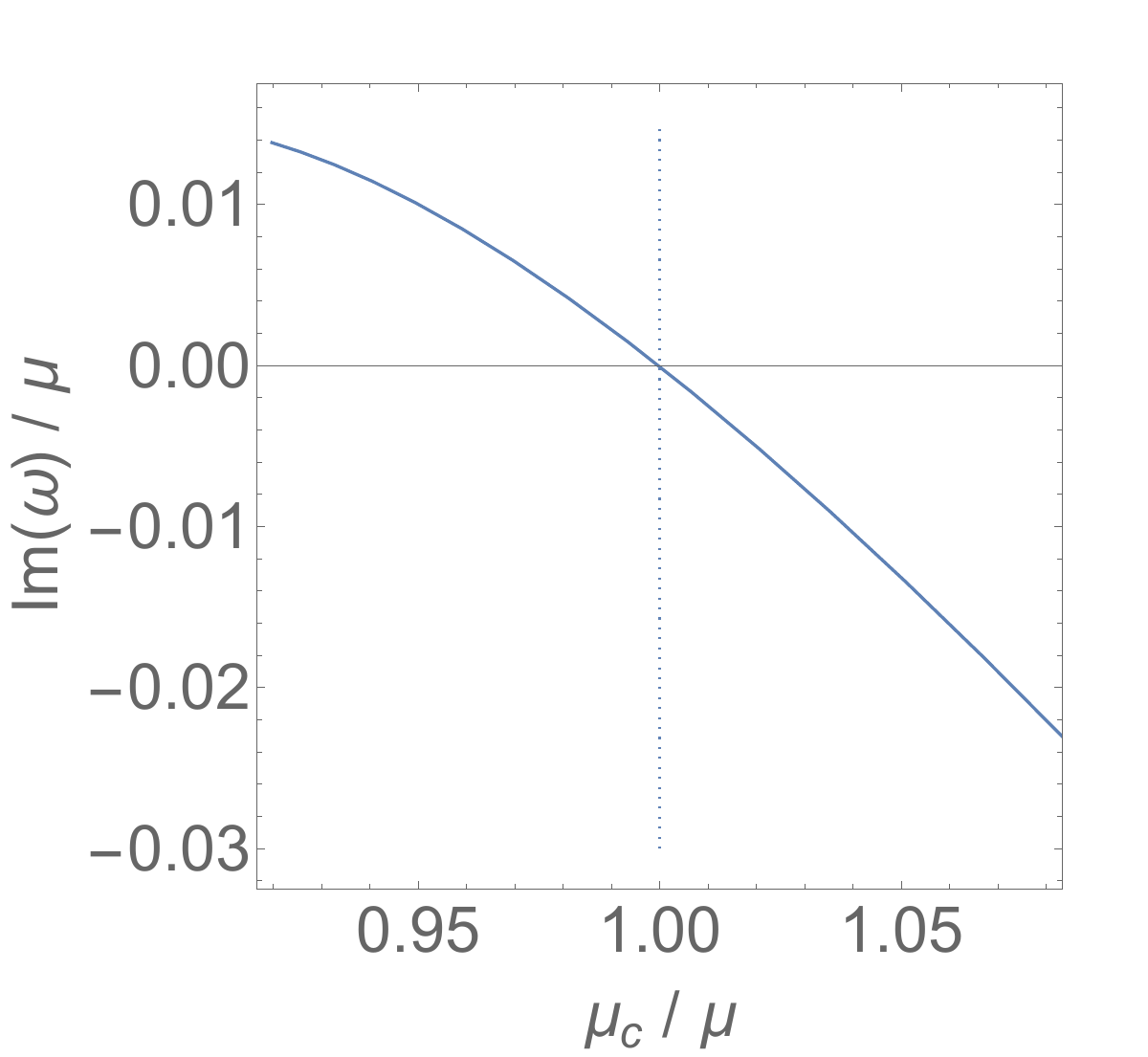}
\caption{\label{decoupledqnms} Dispersion relation of the hydrodynamic mode corresponding to fluctuations of the order parameter $\mathcal{O}$. In the region where $\chi_{\mathcal{O}\mathcal{O}}<0$, shown in figure \ref{decoupledsuperfluidplots}, it is observed that $\Im \omega > 0$, as expected from the analysis in the main text.}
\end{center}
\end{figure}

Following our discussion in the main text, the instability towards condensation of the order parameter should also appear near the phase transition as an unstable hydrodynamic mode with $\Im \omega > 0$. In our small amplitude limit, these can be found via gauge/gravity duality by perturbing the scalar fields at finite frequency,
\begin{align}
\begin{split}
\Psi(r) &= (\sigma_r(r) + i\sigma_i(r))e^{-i\omega t} \,,\\
\Psi^*(r) &= (\sigma_r(r)- i\sigma_i(r))e^{-i\omega t}
\end{split}
\end{align}
with ingoing boundary conditions at the horizon,
\begin{align}
\lim_{r\to 1}\sigma_{r,i} = e^{-i\omega r_*}\sigma_{r,i}^0\,.
\end{align}
Here $dr_* / dr = f^{-1}(r)$ is the tortoise coordinate. Furthermore, we require no source at the boundary
\begin{align}
\lim_{r\to 0} r\sigma_{r,i} = 0.
\end{align}
The equations of motion for $\sigma_r$ and $\sigma_i$ have solutions only at discrete values of $\omega$ with these boundary conditions. These values define the quasinormal modes and those with $|\omega/\mu|\lesssim 1$ correspond to the hydrodynamic modes, or poles of the response functions, for the dual quantum theory \cite{Kovtun:2005ev}. As expected, in figure \ref{decoupledqnms}, we see an instability in the region where $\chi_{\mathcal{O}\mathcal{O}}<0$.

In this Section, all data must be obtained numerically. Figures \ref{decoupledsuperfluidplots} and \ref{probesuperfluidplots} were obtained by solving the associated ordinary differential equations, eq. (\ref{eq:decoupledEOM}) and (\ref{eq:probeEOM}) over a Chebyshev grid with a Newton-Raphson relaxation algorithm. Figure \ref{decoupledqnms} was obtained via discretization of the linearized equations of motion over a Chebyshev grid and using Mathematica's built in direct eigenvalue solver as explained for example in \cite{Dias:2015nua, Davison:2022vqh, Gouteraux:2022qix, Arean:2023nnn}. The region $r \in [1,\infty) $ was mapped to a finite domain by introducing the auxiliary coordinate $z = 1/r$. For all plots, we used $41$ grid points.


   \end{document}